\documentclass[pra,amsmath,amssymb,twocolumn]{revtex4-1}

\usepackage[normalem]{ulem}
\usepackage{amsmath,amssymb}
\usepackage[bookmarks=true,colorlinks,linkcolor=blue,urlcolor=blue,citecolor=blue]{hyperref}
\usepackage{graphicx,amsmath,amsfonts,dsfont}
\usepackage[usenames]{color}
\usepackage{epstopdf}
\usepackage{verbatim}
\usepackage{amsthm}
\usepackage{enumitem}
\usepackage{hyperref}

\newcommand{\be}{\begin{equation}}
\newcommand{\beq}{\begin{eqnarray}}
\newcommand{\eeq}{\end{eqnarray}}

\def \be{\begin{equation}}
\def \ee{\end{equation}}
\def \ba{\begin{array}}
\def \ea{\end{array}}
\def \bea{\begin{eqnarray}}
\def \eea{\end{eqnarray}}

\def \D{{\Delta}}

\newcommand{\cre}[2]{#1^{\dagger}_{#2}}
\newcommand{\cres}[2]{\bar{#1}_{#2}}

\newcommand{\ann}[2]{#1^{\phantom{\dagger}}_{#2}}
\newcommand{\anns}[2]{#1_{#2}}
\newcommand{\eq}[2]{\begin{equation}\label{#1} #2 \end{equation}}
\newcommand{\sub}[1]{_{\mbox{\tiny #1}}}

\numberwithin{equation}{section}

\newcommand{\changed}[1]{{\color{black}#1}}

\begin{document}

\title{Non-Equilibrium Universality in the Heating Dynamics of Interacting Luttinger Liquids}
\author{Michael Buchhold$^{1,2}$ and Sebastian Diehl$^{1,2}$
\\{$^1$\small \em Institute for Theoretical Physics, Technical University of Dresden, 01062, Dresden, Germany}\\
{$^2$\small \em  Institute for Theoretical Physics, University of Innsbruck, A-6020 Innsbruck, Austria}\\
}
\begin{abstract}
We establish a new non-equilibrium scaling regime in the short time evolution of one-dimensional interacting open quantum systems subject to a generic heating mechanism. This dynamical  regime is characterized by uncompensated phonon production and a super-diffusive, universal scaling of  quasiparticle lifetimes with momentum $\sim q^{-5/3}$, distinct from finite and zero temperature cases. It is separated from a high momentum regime by a time dependent scale fading out as $q_0(t) \sim t^{-4/5}$. In the latter region we observe thermalization to an effective time-dependent equilibrium with linearly increasing temperature. By mapping out the dynamical phase diagram and computing the dynamical structure factor within an open system Keldysh functional integral approach, we show how these predictions can be explored in cold atom experiments by means of Bragg spectroscopy. 
 \end{abstract}
\maketitle

\section{Introduction}
Universality -- the insensitivity of long wavelength macroscopic observables to  the microscopic details of a given physical system -- is a powerful concept in equilibrium many-body physics. Particularly low dimensional systems show a strong degree of universality, which is reflected in the low-energy description of both bosonic and fermionic one-dimensional systems in terms of Luttinger liquids \cite{haldane81,giamarchibook}, where microscopic physics enters only via the value of two independent parameters governing the non-interacting Luttinger Hamiltonian. Static equilibrium properties are accurately described in terms of the correlation functions for  this free Hamiltonian. In contrast, as recognized in seminal early work by Andreev, dynamic, finite frequency equilibrium observables, such as the dynamic structure factor, are governed by non-linear effects \cite{andreev80}. In particular, Andreev predicted a universal, super-diffusive scaling of the particle lifetime with momentum $\sim q^{-3/2}$ for finite temperatures (see also \cite{samokhin98,narayan02,lepri03}). This result has been related to the famous Kardar-Parisi-Zhang equation \cite{KPZ} recently \cite{vanBeijeren, lamacraft13}, and put into a domain of validity for ultracold gases in \cite{gangardt13}. Moreover, the zero temperature quantum limit has been shown to exhibit a different universal, diffusive scaling $\sim q^{-2}$ \cite{affleck06,zwerger06, caux06}.

Given the strong notion of universality in equilibrium in one spatial dimension, a key question is whether and in which precise sense this leverages over to non-equilibrium conditions. This is particularly pressing in the light of recent experiments preparing and probing the nature of low entropy  quantum wires \cite{schmiedmayer12,schmiedmayernphys12,bloch13,nagerl13}. From this, but also from a fundamental theoretical perspective, it is highly desirable to identify universal yet directly observable aspects of many-body dynamics, where the notion of insensitivity not only refers to the microscopic details, but also extends to the initial conditions. Beautiful examples of dynamical universality have been identified in the  dynamics of closed, Hamiltonian systems in \cite{berges_pretherm,burkov07,mitragiamarchi11,mitragiamarchi12,Karrasch12, Tavora13,huberaltman11,heyl13,lux13}.

In this work, we address a natural situation in the context of \emph{open} quantum systems: The many-body dynamics of bosons prepared in their ground state, and exposed to a weak, number conserving heating mechanism, as ubiquitous in experiments with ultracold atomic systems. Specifically, we focus on the short time domain of such a system -- which ultimately reaches the infinite temperature state -- where it is well described by a non-linear Luttinger liquid. This setting may be viewed as a continuous counterpart of a quantum quench \cite{cardycalabrese06}, where energy is injected softly but permanently, instead of suddenly. We develop the theoretical framework and a first physical picture of universal aspects in this interacting quantum dynamics.  In particular, we find that the specific nature of the interactions leads to a remarkably simple structure characterized by a decoupling of the forward or ``ageing'' time evolution, and the frequency resolved dynamic properties. The latter have to be treated fully non-perturbatively, while former is captured by a quantum kinetic equation in the self-consistent Born approximation \cite{Marino12}, featuring the non-perturbatively evaluated self-energies at each time step. On this basis we obtain the following key results. (i) \emph{Low momentum non-equilibrium scaling} -- We demonstrate the robust presence of a window of momenta $q$, which is dominated by phonon production and governed by quasiparticle lifetimes which are neither thermal $\sim q^{-3/2}$, nor zero temperature $\sim q^{-2}$, but rather are dictated by a new non-equilibrium super-diffusive scaling law $\sim q^{-5/3}$ in between the known cases. The existence of this regime is granted for low temperature intial states by a combination of particle number conservation and systematic derivative expansion of the gapless problem.\\
(ii) \emph{High momentum effective thermalization} -- The phonon production regime is separated from a scattering dominated region, where we observe thermalization into a quasi-equilibrium with a time-dependent, increasing temperature (see also \cite{schachen14} for a numerical investigation). The crossover momentum scale between both regimes itself satisfies a scaling law $q_0(t) \sim t^{-4/5}\to 0$. It thus ultimately erases the non-equilibrium momentum window in favor of a time-dependent equilibrium state, but delimits the speed of low-frequency thermalization in a power-law fashion.
We determine a  dynamical phase diagram in Fig. \ref{fig:DistDephTemp}. The large extent of the genuine non-equilibrium regime is promising for exploring these results in experiments, and we show that Bragg spectroscopy is a suitable  tool for probing both the universal forward time and the frequency resolved dynamics.

This article is structured as follows. In Sec.~\ref{sec:Mod}, we introduce the underlying microscopic model, the one-dimensional Bose-Hubbard model (BHM) subject to permanent but number conserving heating. We also specify its low energy representation, the heated interacting Luttinger Liquid in a Keldysh path integral framework. In Sec.~\ref{sec:Green}, we discuss the theoretical approach to address the nonequilibrium dynamics in the present system, based on diagrammatic methods, and derive the kinetic equation and self-energy for the elementary phononic excitations of the Luttinger model. Subsequently, we discuss the results obtained within this approach in Sec.~\ref{sec:Res}, with a focus on the scaling solution for the self-energies and the time-dependent phonon density. We conclude in Sec.~\ref{sec:Conc}.

\section{Model} \label{sec:Mod}
We consider the dynamics of bosonic atoms in a one-dimensional optical lattice, as described by the quantum master equation ($\hbar =1$)
\begin{eqnarray}
\partial_t \rho = - i [H, \rho] +\gamma_{\mbox{\tiny E}} \sum_i [2\hat n_i \rho \hat n_i - \{\hat n_i^2,\rho\}],
\end{eqnarray}
where $H = \sum_i [-J \big(b_i^\dag b_{i+1} + \text{h.c.}\big) + \tfrac{U}{2} \hat n_i(\hat n_i -1)]$ is the Bose-Hubbard Hamiltonian with creation (annihilation) operators $b_i^\dag (b_i)$ and $\hat n_i = b_i^\dag b_i$. $J,U$ are the hopping and  interaction constants, respectively. The dissipative dynamics, generated by hermitian Lindblad operators $\hat n_i$, is the leading, generic contribution due to spontaneous emission from the lattice drive laser  \cite{pichler10} with a microscopic heating rate  $\gamma_{\mbox{\tiny E}}$. \changed{This is equivalent to a locally fluctuating chemical potential and it} leads to dephasing and to a linear increase of the system's energy, $\langle H \rangle (t) \sim \gamma_{\mbox{\tiny E}}t$ \cite{pichler10}. We will be interested in a regime of low filling $\rho_0 = \langle \hat n_i\rangle\ll1$, weak interaction and heating, $U,\gamma\sub{E} \ll J$, and an initial state of the system close to the (superfluid) ground state of the Hamiltonian. 
\subsection{Master equation in the Luttinger description}
For the description of the long wavelength, low frequency dynamics we can work in the continuum limit $b_i \to b(x)$, and introduce a standard Luttinger liquid representation of the field operators \cite{haldane81,giamarchibook}, 
\begin{eqnarray}
b(x) &\approx& \sqrt{\rho(x)}e^{i\theta (x)}\label{fieldop}\\
\rho (x) &\approx& \rho_0 + \partial_x\phi(x)/\pi.\end{eqnarray} 
The smooth component of density fluctuations $\phi(x)$ and the phase fluctuations $\theta(x)$ are conjugate variables, $[\partial_x\phi(x) ,\theta(x)] = i\pi \delta(x-x')$. The resulting continuum master equation, valid on length scales larger than $x_c\approx 1/(\sqrt{\rho_0Um})$ \footnote{The effective mass $m^{-1}=\partial_q^2\epsilon_q$ in the lattice evaluates to $m=(4Ja^2)^{-1}$, $a$ being the lattice constant.}, thus reads
\begin{eqnarray}\label{eq:cont}
\partial_t \rho &=& - i [H, \rho] +\tfrac{2\gamma\sub{H}}{K\pi} \int_x [2\partial_x \phi\rho\partial_x \phi - \{(\partial_x \phi)^2,\rho\}],\\\nonumber
H &=&  \tfrac{1}{2\pi}\int_x [\nu K( \partial_x \theta)^2 + \tfrac{\nu}{K}(\partial_x \phi)^2 
 + \kappa_c (\partial_x \phi)(\partial_x \theta)^2 ],
\end{eqnarray}
with an effective heating rate $\gamma\sub{H}$ \footnote{The effective heating rate is defined as ${\gamma\sub{H}=\gamma_{\mbox{\tiny E}}\left(\sum_{q<\Lambda}2\nu q^2\right)^{-1}}$.}.
At weak coupling, the Luttinger parameters are  $\nu =\sqrt{\tfrac{\rho_0U}{m}}$, ${K=\tfrac{\pi}{2}\sqrt{\tfrac{\rho_0}{Um}}}$. We keep the leading non-linearity resulting from the expansion of the quantum pressure term in the effective low energy Hamiltonian ($\kappa_c = 1/m$) \footnote{A term $\sim(\partial_x \phi)^3 $, not present in the microscopic theory, only slightly modifies prefactors, but not the scaling laws, while a contribution $\sim (\partial_x \theta)^3$ is ruled out by the $\theta\to -\theta$ symmetry of the Hamiltonian.}. The non-linearities are irrelevant for the description of any static correlation function of the Luttinger liquid, but indispensable  for capturing quantitatively dynamic correlation functions \cite{andreev80} as well as the forward time dynamics addressed below. The heating term becomes quadratic in the Luttinger representation, and crucially preserves the gapless, collective nature of the Hamiltonian problem. 

The above heating mechanism may seem rather specific to optical lattices. However, the linear increase in system energy is ubiquitously observed, also in experiments in the spatial continuum \cite{trupke13}. This effect is captured by the continuum heating term in Eq.~\eqref{eq:cont}, and we may thus view it as the leading order in a generic model for heating in the long wavelength limit \footnote{\changed{While the present heating mechanism corresponds to a fluctuating chemical potential, i.e. to the class of Langevin equations with additive noise, it is also possible to imagine mechanisms with multiplicative noise, such as fluctuating interaction parameters. However, while these would correspond to a completely different universality class, typically featuring an exponential energy increase, which is not relevant for optical lattice experiments, where the energy increase is linear in time.}}.
The key property of the heating term exploited here is its particle number conserving nature. This guarantees the existence of a hydrodynamic linear sound mode as long as the system is in its low entropy ordered phase, where the Luttinger description is appropriate. This sharp mode in turn underlies the universality established here, as argued below. It is in stark contrast to an open system with particle number exchange, where the low frequency dynamics is diffusive to leading order. Clearly, the permanent heating ultimately leads to a breakdown of the Luttinger description, and the corresponding time scale is determined below. Our analysis concentrates on the preceeding short-time behavior, and is complementary to the late time asymptotics studied in \cite{poletti12,poletti13,caibarthel13}.

\changed{A further comment on the heating is in order at this point. As one can see from Eq.~\eqref{eq:cont}, the effect of the heating term is becoming stronger for shorter wavelengths. Without an appropriate cutoff, this leads to an immediate breakdown of the Luttinger description, since high momentum modes become very strongly populated. However, as has been shown in Ref.~\cite{pichler10}, the population of higher bands due to spontaneous emission is suppressed exponentially strongly, such that
there exists a very natural cutoff, which on a microscopic scale is set by the inverse lattice spacing. We have verified with explicit numerical simulations, that the exact value at which the heating is cut off does not modify the dynamics of the system as long as the rate with which energy is pumped into the system is kept fixed. The non-universal properties in the heating dynamics depend on numerical value of this heating rate (and therefore implicitly on the way in which the heating cutoff is implemented). However, we want to stress that the universal results, which we focus on in this work, are independent of the heating rate and the cutoff and are in this sense indeed universal.}

In order to prepare for a detailed theoretical analysis, we perform a canonical Bogoliubov transformation of both quadratic and cubic terms. That is, we expand the hermitian field operators into physically more transparent phononic creation and annihilation operators $\cre{a}{q}, \ann{a}{q}$ according to 
\begin{eqnarray}
\theta (x) &=& \theta_0 +i\int_q\left(\tfrac{\pi}{2|q|K}\right)^{1/2} e^{-iqx}\left(\cre{a}{q}-\ann{a}{-q}\right),\label{TrafoT}\\
\phi(x)&=&\phi_0-i\int_q\left(\tfrac{\pi K}{2|q|}\right)^{1/2}\mbox{sgn}(q)\ e^{-iqx}\left(\cre{a}{q}+\ann{a}{-q}\right).\label{TrafoP}\ \ \ \ \ \ \ \
\end{eqnarray}
 The master equation in the phonon basis is
\begin{eqnarray}
\partial_t\rho&=&-i\left[H\sub{ph},\rho\right]+\sum_q 2\gamma\sub{H}|q|\left[\left(\cre{a}{q}+\ann{a}{-q}\right)\rho\left(\ann{a}{q}+\cre{a}{-q}\right)\right. \nonumber\\
&-&\left. \tfrac{1}{2}\left\{\left(\ann{a}{q}+\cre{a}{-q}\right)\left(\cre{a}{q}+\ann{a}{-q}\right),\rho\right\}\right]\label{Supp6}
\end{eqnarray}
with the phonon Hamiltonian
\eq{Supp7}{
H\sub{ph}=\sum_q \nu|q|\cre{a}{q}\ann{a}{q}+H^{(3)}\sub{ph}.}
Here, $H^{(3)}\sub{ph}$ contains cubic phonon scattering processes, resulting from the cubic part of the Hamiltonian in Eq.~\eqref{eq:cont}.

\subsubsection{Dynamics from the quadratic part}
The quadratic part of the master equation \eqref{Supp6} describes the heating of linear dispersing phonon modes, which leads to a linear increase of the phonon occupation in time. This is most easily seen by evaluating the time evolution of quadratic operators by neglecting the cubic part of the Hamiltonian. Using the adjoint equation of the master equation \eqref{Supp6}, one derives the Heisenberg equations of motion for the operators 
\eq{Supp8}{
\partial_t \hat{n}_q=\gamma\sub{H}|q|\Rightarrow \hat{n}_q(t)=\hat{n}_q(0)+\gamma\sub{H}|q|t,
}
with $\hat{n}_q=\cre{a}{q}\ann{a}{q}$. For the anomalous operator $\hat m_q:=\cre{a}{q}\cre{a}{-q}$, one finds
\begin{eqnarray}
\partial_t \hat m_q &=&-\gamma\sub{H}|q|-2i\nu|q| \hat m_q \label{Supp9}\\
& \Rightarrow & \hat m_q(t)=i\tfrac{\gamma\sub{H}}{2\nu}\left(e^{-2i\nu|q|t}-1\right)+e^{-2i\nu|q|t}\hat m_q(0).\nonumber
\end{eqnarray}
The linear increase of the phonon number in time, with a momentum dependent rate $\Gamma_q=\gamma\sub{H}|q|$, in turn leads to a linear increase of the system energy in time, consistent with previous results \cite{pichler10}. In contrast, $|m_q(t)|$ is bounded to a very small value and therefore of negligible influence on the dynamics as we briefly discuss later.

At this point, two further comments are in order. \\
(i)  \emph{UV cutoff} -- In order not to pump an infinite amount of energy into the system, the heating has to be cut-off at some ultraviolet (UV) momentum $q_h$. This is an artifact of  taking the continuum limit of the heating Liouvillian in the main text without accounting for the finite width of the lowest Bloch band, for which $\mathcal{L}$ is defined. A similar problem occurs for correlation functions in the Luttinger Liquid theory, which are commonly regularized introducing an exponential cutoff $e^{-|p|\alpha}$ for the creation and annihilation operators\cite{giamarchibook}.
 However, the precise form of the cutoff does not modify the results of our analysis, as long as $q_h$ is sufficiently large to not cause discontinuities in the time evolution, and we therefore set $q_h=\Lambda$, with $\Lambda$ the cutoff of the Luttinger theory. For a given heating rate, the microscopic heating rate $\gamma_E=\partial_{\tau}E(\tau)$ must be independent of the cutoff and determines the effective heating rate $\gamma\sub{H}$ implicitly via $\gamma_E=\gamma\sub{H}\sum_{q<q_h}2\nu q^2$.\\
(ii) \emph{Adequacy of the Luttinger representation} -- The main physical ingredient of the Luttinger representation of the bosonic field operators Eq.~\eqref{fieldop} is the fact that the density fluctuations $\delta \rho(x) = \partial_x\phi(x)$ are gapless, which is due to the collective nature of one-dimensional systems. Intuitively, this collective nature should be preserved in an exactly number conserving system such as the one considered here. Indeed, formally the dissipative term in the master equation with hermitian Lindblad operators $\rho(x) = \rho_0 + \partial_x\phi(x)$ is invariant under a constant shift $\rho(x) \to \rho (x) -\rho_0$, which together with the familiar form of the Hamiltonian demonstrates the gapless, collective nature of the master equation. This underlies the existence of a sharp collective, coherent phonon mode with dynamical exponent $z=1$ and subleading dissipative corrections. This should be contrasted with a number non-conserving system, where a finite density results from a balance of loss and pumping terms, as e.g. described by non-hermitian Lindblad operators such as $b(x),b^\dag(x)$. The dissipative term in a corresponding master equation does not exhibit the above shift invariance, and so its gapless nature is not obvious. In fact, in such a situation no coherent mode exists at long wavelength. Instead, the leading dynamics is dissipative, with a dissipative dynamical exponent governed by the Kardar-Parisi-Zhang universality class \cite{altman13}. This circumstance would completely invalidate the approach taken here. 
\subsubsection{Resonant three-phonon scattering}
Applying the transformation \eqref{TrafoT}, \eqref{TrafoP} to the cubic part of the Hamiltonian \eqref{eq:cont}, leads to the cubic phonon scattering term
\begin{eqnarray}
H^{(3)}\sub{ph}&=&\int_{q,p}\left\{ \tfrac{1}{3}V_{q,p,-p-q}\ann{a}{q}\ann{a}{p}\ann{a}{-q-p}\right. \nonumber\\
&+&\left. V_{q,p,p+q}\cre{a}{p+q}\ann{a}{q}\ann{a}{p}+\mbox{h.c.}\right\}\label{Supp3}
\end{eqnarray}
with the permutation invariant vertex
\begin{eqnarray}
V(k,q,p)&=&\sqrt{|p\cdot k\cdot q|}v(k,q,p),\label{Supp4}\\
v(k,q,p)&=&\sqrt{\tfrac{\kappa_c^2\pi}{2K}}\left\{\tfrac{qk}{|qk|}+\tfrac{pk}{|pk|}+\tfrac{pq}{|pq|}\right\}.\nonumber
\end{eqnarray}
While momentum conservation is guaranteed by the Hamiltonian \eqref{Supp3}, not all of the processes are also energy conserving with respect to $H^{(2)}$. For instance a process in which three phonons are destroyed or created ($\ann{a}{q}\ann{a}{p}\ann{a}{-p-q}$ or its hermitean conjugate) violate energy conservation since there is a positive energy associated to each phonon. In contrast, the process $\cre{a}{p+q}\ann{a}{p}\ann{a}{q}$ can be energy conserving. Since the dispersion is linear ($\epsilon_q=u|q|$), the process is resonant if
\eq{EnCons}{
|p+q|=|p|+|q|.}
Due to the RG-irrelevant nature of the interaction, only those processes can become relevant for the dynamics, which describe phonons interacting with each other for arbitrary long time without dephasing. These are exactly the resonant processes and we will from now on only consider these \cite{andreev80, zwerger06}. For the case of resonant scattering, the function $v(p+q,q,p)$ takes a constant value
\eq{Supp5}{
v_0:=v(1,1,1)=3\kappa_c\sqrt{\tfrac{\pi}{2K}}.}
 Since $\kappa_c$ is only roughly determined by microscopic parameters $\kappa_c\approx\tfrac{\hbar^2}{m}$, the full interaction strength $v_0$ has to be determined by numerics or inferred from experimental data.
The corresponding Hamiltonian is
\eq{ResHam}{
H\sub{res}=v_0\int_{q,p}' \sqrt{|qp(p+q)|}\left(\cre{a}{p+q}\ann{a}{q}\ann{a}{p}+\mbox{h.c.}\right).}
Here, $\int'$ indicates that the integral runs only over momenta for which the integrand describes resonant scattering.
\subsection{Keldysh action}
The scale invariant, gapless nature of the continuum master equation~\eqref{eq:cont} rules out a perturbative treatment of the non-linearities, which could be performed on the level of the master equation \cite{kessler12,hartmann13,koch13}. It is therefore advantageous to map the  master equation into a fully equivalent Keldysh functional integral \cite{dalla13,sieberer13:_dynam_critic_phenom_driven_dissip_system}, which opens up the problem to non-perturbative techniques from many-body physics.

In a Keldysh path integral framework \cite{Altland/Simons,kamenevbook}, the partition function is defined as the functional integral $\mathcal{Z}=\int\mathcal{D}[\cres{a}{q}^c, \anns{a}{q}^c, \cres{a}{q}^q, \anns{a}{q}^q]e^{i\mathcal{S}}$. The microscopic action $\mathcal{S}$ is a functional of the complex classical and quantum fields $\cres{a}{q}^{c/q}, \anns{a}{q}^{c/q}$, which can be derived directly from the markovian master equation \eqref{Supp6} according to the translation table described in Refs.~\cite{dalla13,siebererlong13}.
The Keldysh action $\mathcal{S}=\mathcal{S}_H+\mathcal{S}_D$ is composed of the Hamiltonian contribution
$\mathcal{S}_H$ and the dissipative part $\mathcal{S}_D$, which reflects the Liouvillian.
The Hamiltonian parts of the action, including the nonlinearities, read \cite{buchholdmethod} 
\begin{eqnarray}
\mathcal{S}_H&=&\frac{1}{2\pi}\int_{t,t',p}\left(\cres{a}{p,t}^c,\cres{a}{p,t}^q\right)\left(\begin{array}{cc} 0& D^R_{p,t,t'}\\D^A_{p,t,t'} & D^{K}_{p,t,t'}  \end{array}\right)\left(\begin{array}{c}\anns{a}{p,t'}^{c}\\ \anns{a}{p,t'}^q\end{array}\right)\hspace{0.5cm}\nonumber\\
&&+\frac{v_0}{\sqrt{8}\pi}\int_{p,k,t}' \sqrt{|pk(k+p)|}\ \Big[ 2\cres{a}{k+p,t}^c\anns{a}{k,t}^c\anns{a}{p,t}^q\nonumber\\&&\hspace{1.2cm}+\cres{a}{k+p,t}^q\left(\anns{a}{k,t}^c\anns{a}{p,t}^c+\anns{a}{k,t}^q\anns{a}{p,t}^q\right)+\mbox{h.c.}\Big],\label{Model26}
\end{eqnarray}
with the bare inverse retarded/advanced propagator
\begin{eqnarray}
D^R_{p,t,t'}&=&\delta(t-t')\left(i\partial_{t'}-u|p|+i0^{+}\right), \\
D^A_{p,t,t'}&=&\left(D^R_{p,t,t'}\right)^{\dagger}=\delta(t-t')\left(i\partial_{t'}-u|p|-i0^+\right)\hspace{1cm}
\end{eqnarray}
and the Keldysh component of the inverse propagator
\eq{Model24}{D^K_{p,t,t'}=2i0^+ F(p,t,t').}
The dissipative action, determined by the Liouvillian is
\eq{Supp11}{
\mathcal{S}_D=i\int_{p,t} \gamma\sub{H} |p|\left(\cres{a}{p}^q,\anns{a}{-p}^q\right)\left(\begin{array}{cc}1 & 1 \\ 1 & 1\end{array}\right)\left(\begin{array}{c}\anns{a}{p}^q\\ \cres{a}{-p}^q\end{array}\right)}
and modifies only the quantum-quantum part of the action. It describes permanent phonon production with a momentum dependent rate $\gamma\sub{H}|q|$ in the diagonal phonon channel, as already derived on an operator level in Eq.~\eqref{Supp8}. Clearly, this cannot be compensated by the non-linearities, which can redistribute energy between the phonon modes but not counteract the energy pump. Therefore a time-independent fluctuation dissipation relation  in this system can only be established at $t\rightarrow\infty$ -- when an infinite temperature steady state is reached.

The off-diagonal phonon production, represented by the $\cres{a}{p}^q\cres{a}{-p}^q+\mbox{h.c.}$ terms is however affected by the quadratic Hamiltonian (c.f. Eq.~\eqref{Supp9}). The off-diagonal density is not continuously pumped by the heating but instead shows oscillatory behavior with a maximum  ${|\langle\cre{a}{p}\cre{a}{-p}\rangle|_{\mbox{\tiny max}}=\tfrac{\gamma\sub{H}d}{\nu}\ll 1}$, due to the interplay of quadratic unitary and dissipative dynamics. \changed{Due to the dominant resonant scattering, there is no scattering from the normal phonon mode into the anomalous phonon mode, which would require a resonant scattering vertex that is cubic in creation operators. However, such a cubic vertex is unable to fulfill energy and momentum conservation (for a detailed discussion of the technical details see \cite{buchholdmethod})}. As a result, the kinetic equation for the diagonal elements can be solved independently of the off-diagonal elements. Furthermore the occupation of the off-diagonal correlator remains small even in the presence of interactions, as shown in Eq.~\eqref{Supp9}. Therefore, we neglect the latter and project the dissipation onto diagonal terms.
The corresponding dissipative action is 
\eq{DisAc}{
\mathcal{S}_D=i\gamma\sub{H}\int_{p,t} |p|\ \cres{a}{p}^q\anns{a}{p}^q.}

In this section, we have derived the Keldysh action for the low energy dynamics of interacting lattice bosons subject to dephasing. The dephasing is caused by spontaneous emission processes of the bosonic atoms. The action $\mathcal{S}=\mathcal{S}_H+\mathcal{S}_D$ consists of a Hamiltonian part $\mathcal{S}_H$, describing an interacting Luttinger Liquid with resonant phonon scattering processes, and a dissipative part $\mathcal{S}_D$ representing the dephasing, which leads to a permanent production of phonons in the system. Here, we have justified and already implemented two approximations, namely the {\it resonant approximation}, taking only energy conserving phonon scattering processes into account and leading to the Hamiltonian action \eqref{Model26}, and the {\it "diagonal" approximation}, which neglects the off-diagonal phonon density and results in the dissipative part \eqref{DisAc}.
Both of them will be used in the next section to derive the quantum kinetic equation and phonon self-energy for the present system.

\section{Phonon Green's functions}\label{sec:Green}
We are interested in the full phononic single-particle Green's functions, i.e. the retarded Green's function
\eq{RetGre}{G^R_{p,t,t'}=\left(D^R-\Sigma^R\right)^{-1}_{p,t,t'}}
and the Keldysh Green's function
\eq{KelGre1}{G^K_{p,t,t'}=\left(G^R\circ F- F\circ G^A\right)_{p,t,t'}.}
Here, $\Sigma^{R}$ is the retarded self-energy generated by the resonant phonon interaction and $F_{q,t,t'}$ is the nonequilibrium phonon distribution function depending on the interplay of dissipation and interaction.
\subsection{Nonequilibrium Fluctuation-Dissipation Relation}
For an out-of-equilibrium situation with explicitly broken time translational invariance, it is useful to introduce Wigner coordinates. Accordingly, a two time function $F_{q,t,t'}$ is expressed in terms of relative time $\Delta_t=t-t'$ and forward time $\tau=\frac{t+t'}{2}$. The Wigner transform of $F_{q,t,t'}$ is defined as
\eq{Supp18}{
F_{q,\omega,\tau}=\int d\Delta_t\ e^{i\omega\Delta_t}F_{q,\tau+\Delta_t/2,\tau-\Delta_t/2}.}
In Wigner coordinates, the nonequilibrium fluctuation-dissipation relation (FDR) 
\eq{NonEqFDR}{
\partial_{\tau}F_{q,\omega,\tau}=i\Sigma^K_{q,\omega,\tau}-i\left(\Sigma^R\circ F-F\circ \Sigma^A\right)_{q,\omega,\tau}}
yields the time-evolution of the distribution function in terms of the Keldysh self-energy $\Sigma^K$ and the retarded/advanced self-energies $\Sigma^{R/A}$. In the present case, the Keldysh self-energy (fluctuation contribution)
\eq{Keldself}{
\Sigma^K_{q,\omega,\tau}=-i\gamma\sub{H}|q|+\tilde{\Sigma}^K_{q,\omega,\tau}}
consists of a term $\tilde{\Sigma}^K$ generated solely by the phonon scattering and a term $\gamma\sub{H}|q|$ stemming from the dephasing. The latter drives the system away from equilibrium: It generates a permanent time evolution, which cannot be compensated by the dissipation contribution (term in brackets in Eq.~\eqref{NonEqFDR}) generated entirely by conservative Hamiltonian contribution to the dynamics.

In the following, we will use two distinct approximations in order to simplify this equation and solve for the phonon Green's functions. We will justify these approximations at the end of the section.\\
First, we exploit the fact, that the forward time evolution and the frequency dynamics decouple due to the subleading nature of the interactions. This justifies the Wigner approximation in time and simplifies Eq.~\eqref{NonEqFDR} significantly. In Wigner approximation, the time evolution can be solved first, and independently of the frequency dynamics, yielding a time-dependent single particle distribution function. In a second step, for each instant in time the impact of the non-linearities on the frequency dynamics can be studied in a quasi-stationary state. This can be seen as a ``local time approximation'' in some analogy to a local density approximation in space. Technically speaking, in Wigner approximation, the Wigner transform of a convolution is identical to the product of the Wigner transforms.\\
Second, due to the RG irrelevant interactions, the phonons become dressed, yet still well-defined quasi-particles. This is expressed by the fact that the phonon spectral function
\eq{specfun}{
\mathcal{A}_{q,\omega,\tau}=i\left(G^R-G^A\right)_{q,\omega,\tau}}
is sharply peaked at the bare phonon energy $\omega=\epsilon_q$ with a typical width $\gamma_q\ll \epsilon_q$ much smaller than the energy. Consequently, all the quasi-particle weight is located at $\omega=\epsilon_q$ and the self-energy and distribution function can be evaluated on-shell. The on-shell self-energies can be parametrized as
\eq{OnshS}{
\Sigma^R_{q,\omega=\epsilon_q,\tau}=-i\sigma^R_{q,\tau}, \ \ \ \ \tilde{\Sigma}^K_{q,\omega=\epsilon_q,\tau}=-2i\sigma^K_{q,\tau}.}
Here, $\sigma^{R/K}$ are {\it positive} and {\it real} functions of momentum and forward time \cite{buchholdmethod}.
The on-shell distribution function for well-defined quasi-particles
\eq{OnshF}{
F_{q,\omega=\epsilon_q,\tau}=2n_{q,\tau}+1} is simply the phonon density \cite{kamenevbook}.

Utilizing both Wigner and quasi-particle approximations, the FDR, Eq.~\eqref{NonEqFDR}, simplifies to
\eq{FDR}{
\partial_{\tau}n_{q,\tau}=\frac{\gamma\sub{H}|q|}{2}+\sigma^K_{q,\tau}-\sigma^R_{q,\tau}\left(2n_{q,\tau}+1\right).}
This is the on-shell nonequilibrium FDR for an open interacting Luttinger Liquid, driven by spontaneous emission processes of the microscopic particles. Due to the first term on the r.h.s. of the FDR, the system is driven away from a thermal equilibrium state and is unable to thermalize to a stationary finite temperature state.
\subsection{Self-energies and kinetic equation}
The on-shell self-energy $\sigma^R_{q,\tau}$ and the kinetic equation for Luttinger Liquids with resonant interactions have been derived in a previous work for an isolated system \cite{buchholdmethod}. Here  modifications arise due to openess of the system, expressed by the presence of the factor $\propto \gamma\sub{H}|q|$ in Eq.~\eqref{FDR}.

We consider a zero temperature initial state with $n_{q,\tau=0}=0$. For this case, the vertex correction is exactly zero at $\tau=0$ \cite{buchholdmethod} and the self-consistent Born approximation becomes justified at sufficiently short times. For this situation, the self-consistency equation, obtained by nonequilibrium diagrammatics, for the retarded self-energy is 
\eq{SelfEn18}{
\tilde{\sigma}^R_q= \int_{0<p}\left(\tfrac{\partial_{\tilde{\tau}}n_p}{\tilde{\sigma}^R_p}+2n_p+1\right)\left(\tfrac{qp(q-p)}{\tilde{\sigma}^R_p+\tilde{\sigma}^R_{q-p}}+\tfrac{qp(p+q)}{\tilde{\sigma}^R_p+\tilde{\sigma}^R_{p+q}} \right).}
Here, we have omitted the forward time index and performed the rescaling $\tilde{\tau}=v_0\tau$, $\tilde{\sigma}^R=\sigma^R/v_0$ in order to make this equation independent of microscopic variables. For a given phonon density $n_p$ and its time derivative $\partial_{\tilde{\tau}}n_p$, it can be solved self-consistently either by using numerics or by a scaling ansatz. The latter is applicable only for the special case for which the phonon density itself is a scaling solution, as we discuss below. In order to obtain the time dependent phonon density a kinetic equation approach is used.

\begin{figure}[t]
  \includegraphics[width=8.6cm]{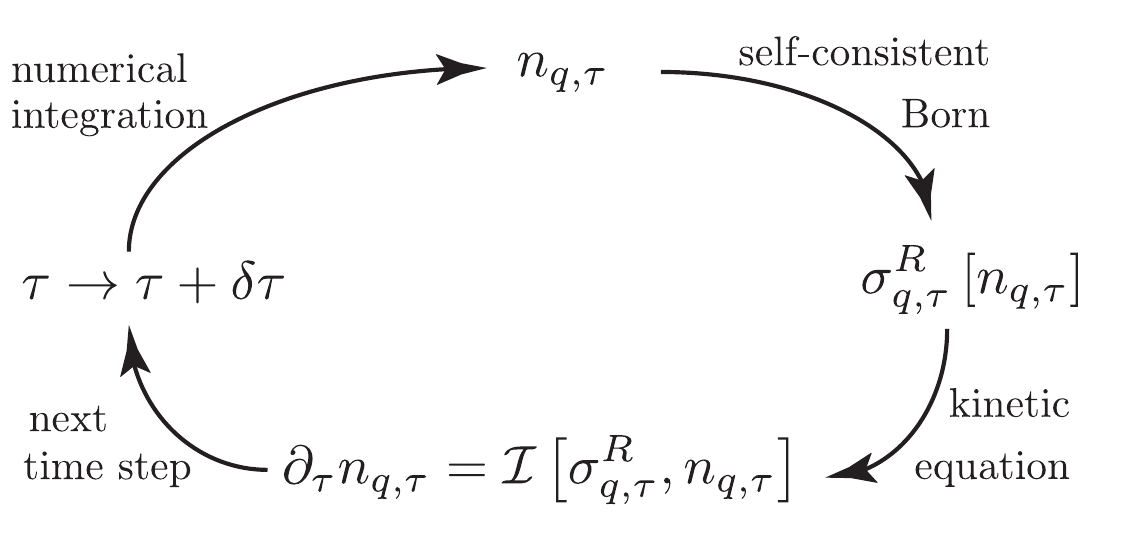}
  \caption{Schematic illustration of the iteration process to determine the time-dependent phonon density $n_{q,\tau}$. For a given time $\tau$, the self-energy $\sigma^R_{q,\tau}$ is determined via the self-consistent Born approximation according to Eq.~\eqref{SelfEn18}. Subsequently the time derivative of $n_{q,\tau}$ is computed via the kinetic equation \eqref{Kinetic5}. Using a Runge-Kutta solver for numerical differential equations, the density $n_{q,\tau+\delta\tau}$ is computed and used as the starting point for the next iteration.}
  \label{fig:Iter}
\end{figure}

The kinetic equation for the phonon density is again derived via the diagrammatic approach, outlined in Ref.~\cite{buchholdmethod}. In terms of the rescaled time $\tilde{\tau}$ and self-energy $\tilde{\sigma}^R$ it reads
\begin{eqnarray}
\partial_{\tilde{\tau}}n_q\hspace{-1mm}&=&\frac{\gamma\sub{H}|q|}{2v_0}\label{Kinetic5}\\
&&+\hspace{-0.5mm}\int_{0< p < q}\hspace{-0.7cm}\frac{2pq(q-p)\left(n_pn_{q-p}-n_q\left(1+n_p+n_{q-p}\right)\right)}{\tilde{\sigma}^R_q+\tilde{\sigma}^R_p+\tilde{\sigma}^R_{q-p}}\ \ \nonumber\\
&&+\hspace{-0.5mm}\int_{0<p}\hspace{-4mm}\frac{4pq(q+p)\left(n_{p+q}\left(n_q+n_p+1\right)-n_qn_p    \right)}{\tilde{\sigma}^R_q+\tilde{\sigma}^R_p+\tilde{\sigma}^R_{q+p}}.\nonumber
\end{eqnarray}
The kinetic equation determines the time evolution of the phonon density in the system $n_{q,\tau}$. Together with Eq.~\eqref{SelfEn18} it forms a closed set of equations for the nonequilibrium dynamics of the driven, interacting Luttinger Liquid. Both can be solved iteratively according to the scheme depicted in Fig.~\ref{fig:Iter}.

A specific but relevant case is the kinetic equation for small momenta $q\ll 1$. For this condition fulfilled, it simplifies to
\eq{KinEqSmall}{
\partial_{\tilde{\tau}}n_q\overset{q\ll 1}{=}|q|\left(\frac{\gamma\sub{H}}{2}+\mathcal{I}_{\tilde{\tau}}\right),}
where 
\eq{IIn}{
\mathcal{I}_{\tilde{\tau}}=\int_{0<p}\frac{2p^2\left(n_{p,\tilde{\tau}}+1\right)n_{p,\tilde{\tau}}}{\tilde{\sigma}^R_{p,\tilde{\tau}}}}
is a time dependent but momentum independent function. Consequently, the change of $n_q$ is linear in the momentum $q$ for small momenta.

One should note, that except for the initial phonon distribution $n_{q,\tau=0}$ and the heating term $\propto\gamma\sub{H}$, no additional microscopic information enters the dynamics expressed in the kinetic equation \eqref{Kinetic5} and self-energy equation \eqref{SelfEn18}. In particular, both equations are insensitive to an UV cutoff for the Luttinger Liquid $\sim\sqrt{U\rho_0 m}$ for sufficiently strongly decaying distribution $n_{p,\tilde{\tau}}$, which is provided, e.g., for zero or finite temperature initial states. Consequently, the dynamics induced by the heating is universal, in the sense that it only depends on the energy pump into the system $\propto\gamma\sub{H}$ and the initial state.  We will now close this section with a discussion of the validity of the above made approximations, namely the Wigner and the quasi-particle approximation.

\subsection{Validity}
In the previous sections, we have used two essential approximations, namely the Wigner approximation and the quasiparticle approximation, in order to obtain analytical results for the self-energy and kinetic equation of the phonons. Although these are quite standard approximations in the kinetic theory of interacting particles, they have to be justified properly. In this section, we will show that both approximations are valid in the present case and give a proper bound for their applicability.
\subsubsection{Wigner approximation }
\changed{We applied the Wigner approximation in order to simplify the Wigner transform of the convolution in the FDR \eqref{NonEqFDR} and approximate it by the lowest order contribution. As a consequence, the Wigner  transform of the product in \eqref{NonEqFDR} equals the product of the individual Wigner transformed functions, i.e.
\begin{eqnarray}
\left(\Sigma^R\circ F\right)_{q,\omega,\tau}&=&\Sigma^R_{q,\omega,\tau}e^{\frac{i}{2}\left(\overset{\leftarrow}{\partial}_{\tau}\overset{\rightarrow}{\partial}_{\omega}-\overset{\leftarrow}{\partial}_{\omega}\overset{\rightarrow}{\partial}_{\tau}\right)}F_{q,\omega,\tau}\nonumber\\
&\approx&\Sigma^R_{q,\omega,\tau}F_{q,\omega,\tau}\label{Wignerapp}.
\end{eqnarray}
As one can see, in the Wigner approximation, all the terms beyond zeroth order in the expansion of the exponential are neglected. This is justified, if the contribution of these terms, which contains derivatives of the self-energy and the distribution function, is negligibly small.
We will now show that the time scale for the breakdown of the Wigner approximation lies on the order of the time scale where the Luttinger liquid description breaks down, and thus never poses an additional restriction to our approach. 

The condition for which the Wigner approximation in Eq.~\eqref{Wignerapp} is justified reads
\eq{Supp20}{
1\gg \underbrace{\left|\frac{\partial_{\omega}\Sigma^R_{q,\omega,\tau}}{\Sigma^R_{q,\omega,\tau}}\right|}_{t_{\Delta_t}}\underbrace{\left|\frac{\partial_{\tau}n_{q,\tau}}{n_{q,\tau}+\tfrac{1}{2}}\right|}_{1/t_{\tau}},}
which is equivalent to requesting the first order term of the expansion of the exponential to be already much smaller than the zeroth order term.}
Physically, this means that the characteristic time scale of the forward dynamics $t_{\tau}$ is much larger than for the relative dynamics. The characteristic time scale for the relative dynamics is determined by the Hamiltonian time evolution, i.e. is set by the inverse quasi-particle energy, while the characteristic forward time scale is set by the dissipation on the r.h.s of Eq.~\eqref{Supp20}, which is generated by the subleading interactions. 
As we show now, the criterion \eqref{Supp20} is fulfilled for sufficiently small $\gamma\sub{H}\tau$, with heating rate $\gamma\sub{H}$ and forward time $\tau$.
To demonstrate this, we use the result for the phonon lifetimes from the subsequent section and evaluate it on shell (i.e. for $\omega=\nu |q|$) to find ${t_{\Delta_t}=\left|\frac{\partial_{\omega}\Sigma^R_{q,\omega,\tau}}{\Sigma^R_{q,\omega,\tau}}\right|=\frac{1}{\nu}\left|\frac{\partial_{q}\sigma^R_{q,\tau}}{\sigma^R_{q,\tau}}\right|=\tfrac{\eta_R}{\epsilon_q}}$ with $\eta_R = \mathcal O (1)$ as seen below. To estimate the time-scale of the forward evolution, we use the decay integrals in \eqref{Kinetic5} and evaluate them in the thermalized regime (this sets a lower bound: in the low momentum regime, the occupation is much lower than a corresponding thermal occupation, leading to a slower decay compared to the thermal regime, and the Wigner approximation is valid for even longer times) to get
\eq{Supp25}{
\frac{1}{t_{\tau}}=c_0\frac{v_0^2}{\nu^2}T(\tau)q^2,}
with $c_0=\mathcal{O}(1)$. The increase in total energy is only caused by the heating term and is linear in $\gamma\sub{H}\tau$. As a consequence, we find $T(\tau)=T_0 \gamma\sub{H}\tau$ with $T_0=\mathcal{O}(\tfrac{1}{10})$ typically. Evaluating \eqref{Supp20} at the UV cutoff momentum again in a conservative way, $q=\Lambda$ (with $\Lambda\approx \nu/v_0$, $v_0$ defined in Eq.~\eqref{Supp5}) yields
\eq{Supp27}{
\tau\ll \frac{1}{c_0\eta_R}\frac{\nu^2}{T_0  v_0\gamma\sub{H}}.}
Comparing this to the criterion for the validity of the Luttinger Liquid description $T(\tau)\ll\nu \Lambda$, which we can rearrange to
\eq{Supp28}{
\tau\ll\frac{\nu^2}{T_0 v_0\gamma\sub{H}},
}
shows that the Wigner approximation is justified for times $\tau$, for which the Luttinger description is applicable.

\subsubsection{Quasi-particle approximation}

In the quasiparticle approximation, the elementary excitations (in the present case the phonons) are dressed but well defined quasiparticles in the sense that their spectral weight is essentially located at a specific frequency $\omega=\epsilon_q$, the quasiparticle energy. Consequently, physical quantities like the self-energy and the distribution function can be evaluated on-shell (at $\omega=\epsilon_q$).  As already stated above, this requires ${\epsilon_q\gg\left|\mbox{Im}\left(\Sigma^R_{q,\omega=\epsilon_q,\tau}\right)\right|}$. In other words, the imaginary part of the self-energy must be much smaller than the quasiparticle energy. For the present case, the on-shell self-energy is purely imaginary and the above condition transforms into
\eq{QPCrit}{
\epsilon_q\gg \sigma^R_{q,\tau}.}

From the time-dependent phonon density $n_{q,\tau}$, we determine below the self-energy $\sigma^R_{q,\tau}$ according to Eq.~\eqref{SelfEn18}. Indeed $\sigma^R_{q,\tau}\ll\epsilon_q$ for all times and momenta considered. Furthermore
\eq{QPCrit2}{\frac{\sigma^R_{q,\tau}}{\epsilon_q}\overset{q\rightarrow0}{\longrightarrow}0} for all times is guaranteed by the subleading nature of the interactions.
Thus, obedience of the quasi-particle criterion is justified \emph{a posteriori}. For the specific case of an equilibrium state (for $T\ge 0$), the quasiparticle approximation breaks down exactly at the UV cutoff $q=\Lambda$.

\section{Time evolution and nonequilibrium scaling}\label{sec:Res}
In this section, we discuss the nonequilibrium dynamics of a heated interacting Luttinger Liquid obtained by solving Eqs.~\eqref{Kinetic5} and \eqref{SelfEn18} numerically as described in the previous section. The main results are the time evolution of the phonon density $n_{q,\tau}$ and a scaling solution for the quasiparticle lifetimes $t_{q}\sim q^{\eta_R}$ with a new nonequilibrium exponent $\eta_R=\frac{5}{3}$. The latter is observable in the low momentum regime for momenta $q<q_0(\tau)$, where $q_0(\tau)$ is a time-dependent momentum scale that separates the nonequilibrium low momentum regime from a quasi-thermal large momentum regime and fades out as $q_0(\tau)\sim \tau^{-\frac{4}{5}}$.

\subsection{Time-dependent phonon density}
\begin{figure}
  \includegraphics[width=8.5cm]{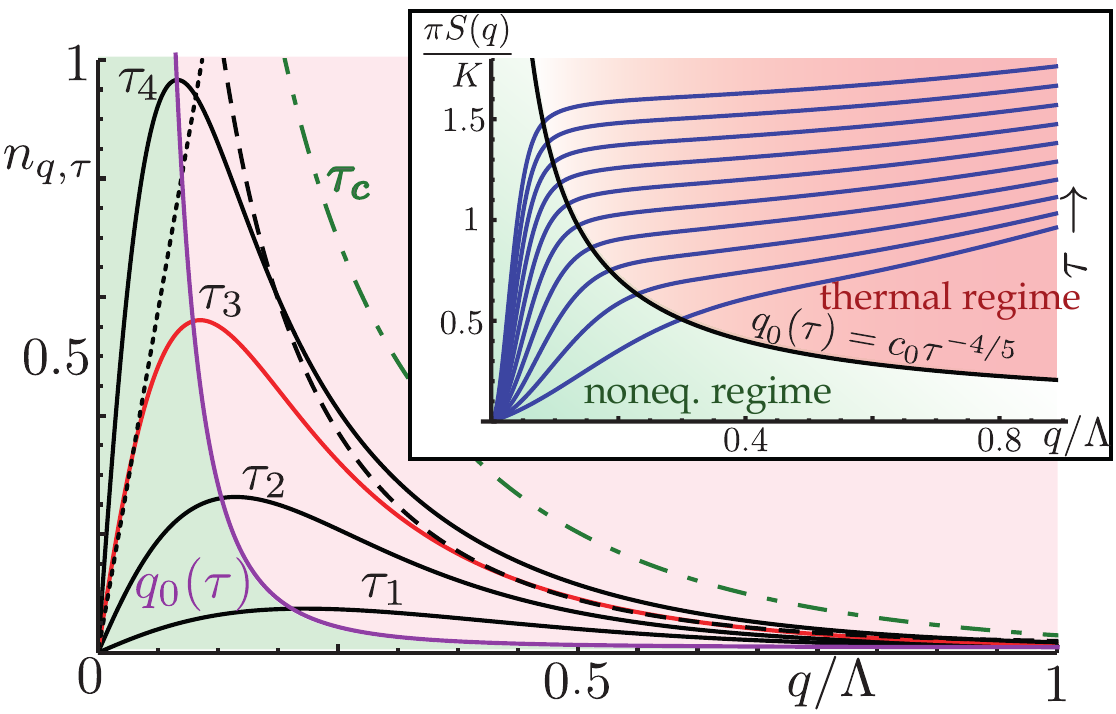}
  \caption{(Color online) Dynamics of the phonon occupation $n_q(\tau)$ in a sequence of times $(\tau_1,\tau_2,\tau_3,\tau_4)=(1,2,3,4)\cdot\frac{10}{v_0\Lambda^2}$ as a function of $q/\Lambda$, with 
  heating rate $\gamma\sub{H}=0.06 v_0\Lambda$. ($\Lambda=1/x_c$ is the high momentum cutoff delimiting the validity of the Luttinger liquid description. For weak interactions $\Lambda \approx \sqrt{\rho_0 m U}$.) Non-equilibrium ($n_q \sim |q|$) and time-dependent thermalized  ($n_q \sim n_B(\nu|q|)$) regimes are indicated by the dotted (dashed) line for $\tau_3$ ($T(\tau_3)=0.38\nu\Lambda$), and separated by the crossover scale $q_0(\tau)$. The dash-dotted line shows $n_B(T(\tau_c))$, where $T(\tau_c)\approx\nu\Lambda$, at which the Luttinger description breaks down.  Inset: Dynamical phase diagram mapped out by the experimentally accessible static structure factor, which is a direct measure of the phonon distribution (see text).
  }
  \label{fig:DistDephTemp}
\end{figure}

The time evolution of the phonon occupation obtained numerically is plotted for different times in Fig. \ref{fig:DistDephTemp}. We clearly identify a (time dependent) crossover scale $q_0(\tau)$, which we find analytically to scale to zero as $q_0(\tau) \sim \tau^{-4/5}$, consistent with numerics. The crossover scale separates a non-equilibrium low momentum regime $n_q(\tau)\sim |q|$ from a scattering dominated thermalized high momentum region $n_q \sim 1/|q|$. 

In the latter regime, the integral and dephasing contributions scale $\sim q^2, |q|$ respectively. As a consequence of this scaling, for sufficiently large momenta, the collision term yields the dominant contribution to the dynamics compared to the heating term. 
The dominant relaxational dynamics for the distribution function then approaches the Bose distribution function $ n_B(c|q|/T(t))$, which is the dynamical fixed point for the collisional term  in Eq.~\eqref{Kinetic5} alone, and where there is an approximate detailed balance between phonon emission and absorption.  Indeed, the occupation number is fitted well with a thermal distribution $n_q = n_B(c|q|,T(t)) \approx T(t)/c|q|$ in this regime. \changed{In a stochastic wavefunction interpretation of the underlying master equation, in this regime multiple thermalizing collisions happen in between two subsequent spontaneous emission events. While the individual heating processes continuously create additional phonons, the system finds the time to relax to a local equilibrium in between two subsequent heating events. Consequently, in the large momentum regime, the heating is described by a time dependent temperature. Such a behavior has been observed numerically in} \cite{schachen14}.

In contrast to this, the low momentum regime is strongly non-thermal; this fact underlies the new scaling law for the particle decay width established below. Here we encounter a strongly overweighing phonon production both due to heating and phonon scattering.  Two structural properties ensure the linear rise with no offset, $n_q(\tau)\sim |q|$. First,  both terms in Eq.~\eqref{Kinetic5} contribute $\sim |q|$ for $|q| \to 0$ as emphasized in the discussion of the low momentum version of the kinetic equation \eqref{KinEqSmall}. This dependence is guaranteed by the form of both dephasing term and scattering vertex, which in turn are dictated by the collective nature of the system, or technically by the systematic derivative expansion of the low frequency action to leading order.\\ Second, the value of $n_{q=0}(\tau)$ at zero momentum is pinned to its initial value for all times. This is a key structural property of the problem, related to particle number conservation: The static structure factor at zero momentum is related to the particle number variance via $S_\tau(q=0) = (\langle \hat N^2(\tau)\rangle - \langle \hat N(\tau)\rangle^2 )/L = S_{\tau=0}(q=0)$, where $\hat N(\tau)$ is the total particle number operator in the Heisenberg picture and $L$ the system length, cf. \cite{mahan}. The last equality then holds in a particle number conserving system as the present one. Since in addition $n_q(\tau) \sim S_\tau (q)$ in the Luttinger  model, cf. Eq. \eqref{eq:staticstruct}, this ensures a  permanent ``pinning'' of the phonon occupation at zero momentum. For zero temperature initial states, these two facts guarantee the existence of a linear phonon occupation regime described above. Remarkably, this mechanism leads to very slow, algebraic thermalization of the low momenta: Establishment of the $1/c|q|$ Rayleigh-Jeans divergence can only be achieved due to the scaling of the crossover momentum $q_0(t)\to 0$, but not via a direct filling of the low momentum modes. We conjecture that this is generic for the thermalization of one-dimensional quantum systems in phases with superfluid order, comprising the setting of a quantum quench.

\subsection{Quasiparticle lifetime scaling and dynamical phase diagram }
We now address 
the frequency dynamics at each time step. In particular, we are interested in the scaling behavior of the phonon quasi-particle lifetimes, which results from the overlap of the coherent phonon mode with the many-body continuum. This is achieved within the Keldysh framework, where the distribution function is not fixed \emph{a priori} to its equilibrium shape, but is replaced by the dynamically determined non-equilibrium distribution.  
The quasiparticle lifetimes $t_q$ are defined via
\eq{QPL}{
t_{q,\tau}=-\left(\mbox{Im}\Sigma^R_{q,\omega=\epsilon_q,\tau}\right)^{-1}=\left(\sigma^R_{q,\tau}\right)^{-1}.} 

As we have seen in the previous subsection, the phonon density can be described by a scaling ansatz in the two different regimes $q\ll q_0$ and $q\gg q_0$. For small momenta, $n_{q,\tau}=|q| f_{\tau}$, where $f$ is a positive function of the forward time, while for large momenta $n_{q,\tau}=\frac{T_{\tau}}{\nu|q|}$ with an effective time-dependent temperature $T_{\tau}$.
From the form of the self-energy equation \eqref{SelfEn18}, we infer directly that if the first factor under the integral  can be written as a scaling form \eq{ScalingForm}{
\left(\left(\tilde{\sigma}^R_{p}\right)^{-1}\partial_{\tilde{\tau}}n_p+2n_p+1\right)= a_{\tau}p^{\eta_n}} with a positive (time-dependent) prefactor $a_{\tau}$, the self-energy itself will be a scaling function (see Ref.~\cite{buchholdmethod} for a detailed analysis of possible scaling forms). In the above mentioned momentum regimes, Eq.~\eqref{ScalingForm} is expected to hold since the phonon occupation $n_p$ has a scaling form, which directly translates to a scaling of the left hand side of Eq.~\eqref{ScalingForm}. Consequently, we parametrize $\tilde{\sigma}^R_q=\gamma_{\tau}|q|^{\eta_R}$.

The scaling equation for the retarded self-energy reads
\begin{eqnarray}
\sigma^R_q&=&\gamma_{\tau} |q|^{\eta_R}=\frac{a_{\tau} |q|^{4+\eta_n-\eta_R}}{\gamma_{\tau}}I_\eta, \\\nonumber
I_\eta &=&\int_0^{\infty}\frac{dx}{2\pi}\left(\frac{|x|^{\eta_n+1}(1-x)}{|x-1|^{\eta_R}+|x|^{\eta_R}}+ (x\to -x)\right),
\end{eqnarray}
with a dimensionless integral whose value depends parametrically on the scaling exponents but otherwise is a pure number. It leads to the scaling relation for the phonon lifetime and occupation exponents $\eta_R,\eta_n$
\begin{eqnarray}\label{eq:scaling}
\eta_R = 2+\frac{\eta_n}{2},
\end{eqnarray}
which is a key result of this work. $\eta_n$ is determined by the solution of the generalized kinetic equation discussed previously. 

Below the crossover scale $q_0$, the phonon density scales linear in momentum $n_q\sim|q|$. Since $\tilde{\sigma}^R_q\sim |q|^{\eta_R}$ with $\eta_R>1$, this means that $\frac{\partial_{\tilde{\tau}}n_q}{\tilde{\sigma}^R_q\left(2n_q+1\right)}\overset{q\rightarrow0}{\longrightarrow}\infty$ and Eq.~\eqref{ScalingForm} is dominated by the first term in brackets. Consequently $\eta_n=1-\eta_R$, which  implies a super-diffusive behavior $\eta_R = 5/3$ with no equilibrium counterpart. 

In contrast, above the scale, $n_q\sim|q|^{-1}$ and $\frac{\partial_{\tilde{\tau}}n_q}{\tilde{\sigma}^R_q\left(2n_q+1\right)}\overset{q\rightarrow\infty}{\longrightarrow}0$ and Eq.~\eqref{ScalingForm} is dominated by the phonon density $\left(2n_q+1\right)\sim|q|^{-1}$ (or $\sim|q|^0=1$, for very large momenta). This leads to $\eta_n=-1$ and as a consequence $\eta_R=\frac{3}{2}$, as it is know for the finite temperature equilibrium case. In contrast, for the largest momenta $\eta_n=0$ and $\eta_R=2$, implying the zero temperature diffusive behavior \cite{affleck06,zwerger06}. 

\begin{figure}
  \includegraphics[width=8.7cm]{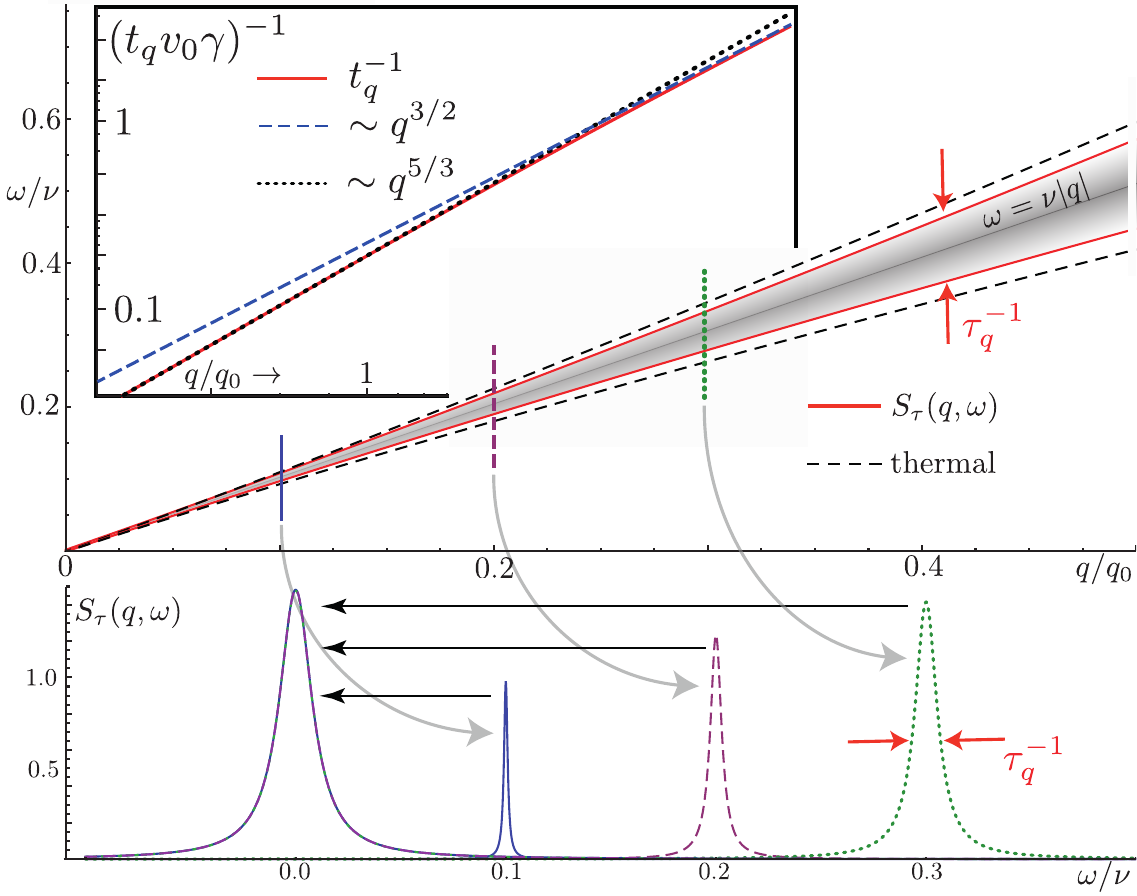}
  \caption{(Color online) Dynamical structure factor (DSF). Upper panel: The DSF  reveals the universal scaling behavior of the phonon lifetimes and allows us to discriminate the new non-equilibrium scaling from equilibrium thermal scaling, both of which are clearly separated. Inset: Log-Log comparison of $t_q$ from $n_q(\tau)$ with the analytically predicted scaling solutions. For small momenta (in units of the crossover scale $q_0$), one clearly identifies the non-equilibrium scaling behavior and only for momenta $q\gg q_0$ the thermal scaling sets in. Lower panel: DSF for fixed momenta $q=(0.1, 0.2, 0.3)$. The plots collapse on top of each other after rescalings $\omega\rightarrow |q|^{-5/3}\big(\omega-\nu|q|\big)$ and  $S\rightarrow |q|^{-1/3}S$,  which demonstrates the scaling $t_q\sim |q|^{-5/3}$.}
  \label{fig:DSF}
\end{figure}

Following the above discussion, we identify the crossover momentum $q_0$ between nonequilibrium and effective equilibrium scaling to be determined by the condition 
\eq{EqNonEq}{
\frac{\partial_{\tilde{\tau}}n_{q_0}}{2n_{q_0}+1}=\tilde{\sigma}^R_{q_0}.}
The scaling behavior of this equation in momentum and time yields
\eq{crossoverSc}{
q_0(\tau)\sim \left(\tau \gamma_{\tau}\right)^{-\frac{1}{\eta_R}}.}
Approaching the crossover from the low momentum regime yields (remember $n_{q,\tau}=f_{\tau}|q|$ for small momenta) $\eta_R=5/3$, $\gamma_{\tau}\sim \left(\frac{f_{\tau}}{\tau}\right)^{\frac{1}{3}}$. Using the numerical result $f_{\tau}\sim\tau^2$, Eq.~\eqref{crossoverSc} can be solved semi-analytically and yields $q_0(\tau)\sim\tau^{-\frac{4}{5}}$. This agrees perfectly with our numerical result for the crossover scale $q_0\sim \tau^{-0.808}$. The semi-analytical analysis in this way demonstrates, that the low momentum regime is responsible for the observed scaling of $q_0$.

\subsection{Experimental detection} We demonstrate that Bragg spectroscopy gives immediate access to all the characteristics of the universal heating dynamics. It enables a direct probe of the density-density correlations in ultracold atomic gases \cite{ketterle99, ketterleI99,  esslinger04, sengstock11}.  More precisely, in a stationary state, the Bragg signal is directly proportional to the Fourier transform of two-point density-density correlations, or dynamical structure factor (DSF) $S(\omega, q) = \int dt dx e^{i (qx - \omega t)}\langle \{ \hat n (t,x),\hat n (0,0)\}\rangle$. In the Keldysh formalism and  long wavelength  limit, this translates into $S_{\tau}(\omega,q ) = - \langle\rho_c(\tau,-\omega,-q)\rho_c(\tau,\omega,q)\rangle$, for which we obtain the explicit expression
\begin{eqnarray}
\hspace{-0.3cm}S_\tau(\omega,q )=\tfrac{(2n_q(\tau)+1)|q|K}{\pi\Sigma^R_q}\big(\tilde{f}\left(\tfrac{\omega-\epsilon_q}{\Sigma^R_q}\right) +  {\scriptstyle{(\omega\to -\omega)\big)}},\ \ \ \
\end{eqnarray}
where $\tilde f (x) = 1/(1+x^2)$ is the dimensionless Lorentzian. As per the above discussion, time enters only parametrically, giving a snapshot, effectively stationary DSF. The different scaling regimes of the quasiparticle lifetimes are clearly identified in Fig.~\ref{fig:DSF} (a). 

The time-dependent crossover scale $q_0(\tau)$  obtains most directly from the equal-time DSF, or static structure factor, 
\begin{eqnarray}\label{eq:staticstruct}
S^s_\tau(q) = \int\frac{d\omega}{2\pi} S_\tau(\omega,q)= \tfrac{|q|K}{\pi}(2n_q(\tau)+1), 
\end{eqnarray}
which is plotted in Fig.~\ref{fig:DistDephTemp}. In particular, it provides direct access to the phonon occupation number, and thus to the scaling of the crossover scale $q_0(\tau)$ and the peak height. 

\subsection{Modifications for $T>0$}
The computations and results presented in the previous sections have been obtained for a system initialized in the $T=0$ ground state. Here, we discuss how an initial finite temperature ($T>0$) state modifies these results, corroborating that the new scaling regime remains observable in an experimentally accessible window of parameters $T_\text{in}\ll q_0(\tau)$. This analysis is relevant and necessary since experiments are never carried out for systems with zero initial but usually very small temperatures. For the Luttinger description to be applicable, the initial temperature must be much lower than the Luttinger ultraviolet cutoff, i.e. $T\ll \nu\Lambda$, $k\sub{B}=1$, which we consider in the following.

\begin{figure}
  \includegraphics[width=8.8cm]{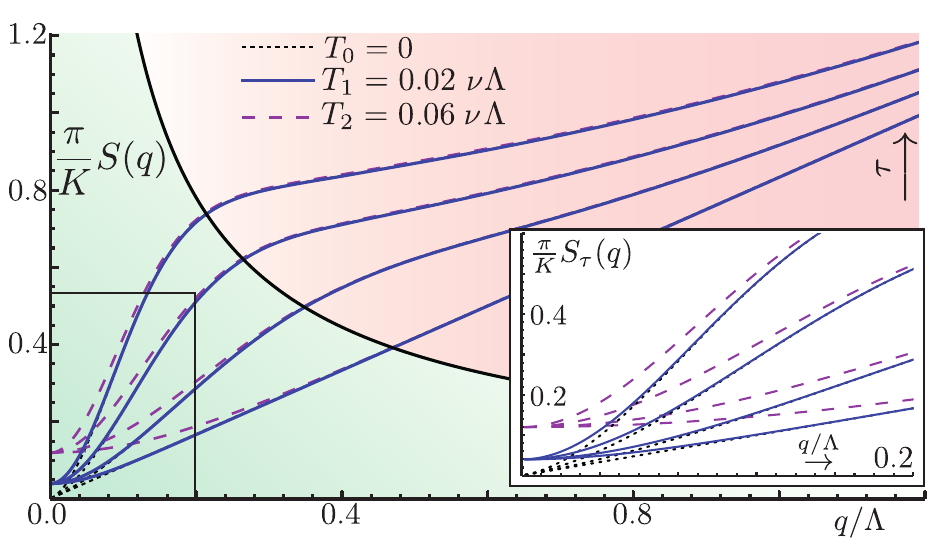}
  \caption{(Color online) Comparison of the time evolved static structure factor $S^s_{\tau}(q)$ for different initial temperatures $T$ at different times $\tau$. The black dotted (blue solid, red dashed) lines corresponds to an initial temperature $T=0$ ($T=0.02\nu\Lambda, T=0.06\nu\Lambda$). The different lines correspond to increasing  times $(\tau_0,\tau_1,\tau_2,\tau_3)=(0, 1, 2, 3)\tfrac{4}{v_0\Lambda^2}$. As for the zero temperature case, $S^s_{\tau}(q=0)$ is pinned to its initial value $S^s_{\tau=0}(q=0)=\tfrac{KT}{\pi}$ due to particle number conserving dynamics. For $0<q<q_0$, the characteristic $S^s(q)\sim q^2$ scaling in the nonequilibrium regime is clearly visible and the $T>0$ curves approach the one for $T=0$. For larger momenta, the finite temperature curves lie on top of the $T=0$ ones. Inset: Zoom into the low momentum region.}
  \label{fig:SSFT}
\end{figure}
\begin{figure}
  \includegraphics[width=8cm]{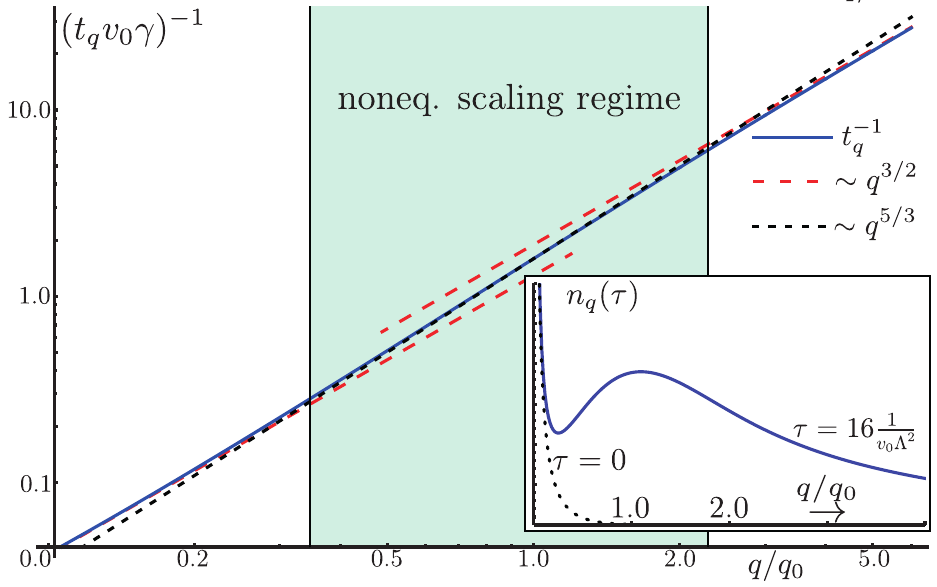}
  \caption{(Color online) Scaling behavior of the inverse quasi-particle lifetime $t_q^{-1}$ for system dynamics initialized at a finite temperature $T=0.02\nu\Lambda$ and evaluated at  time $\tau=\tfrac{16}{v_0\Lambda^2}$. Due to the initial finite temperature, we identify three scaling regimes. For momenta smaller than the initial temperature $q<q_T=\frac{2T}{\nu}$, the $\frac{1}{q}$ divergence of the phonon-distribution dictates a thermal $t_q\sim q^{-3/2}$ scaling behavior. For momenta $q_T<q<q_0$ in between the temperature and the nonequilibrium crossover scale $q_0$, the nonequilibrium scaling $t_q\sim q^{-5/3}$ can be identified, while for $q_0<q$ the system obeys thermal scaling again. As a result, for $q_T\ll q_0$, one will be able to find a sufficiently large scaling regime with the nonequilibrium scaling behavior. Inset: Corresponding phonon distribution $n_q(\tau)$ for times $\tau=0$ (black dotted) and $\tau=\tfrac{16}{v_0\Lambda^2}$ (blue solid) in units of the crossover momentum.}
  \label{fig:ScalingT}
\end{figure}

In order to obtain the time-dependent phonon occupation for the case of an initial $T>0$ state, we numerically solve the kinetic equation, Eq.~\eqref{Kinetic5}, with an initial phonon distribution function
\eq{EqFinT1}{
n_q(\tau=0)=n\sub{B}(\nu|q|)=\left(e^{\frac{\nu |q|}{T}}-1\right)^{-1}.}
However, for a finite temperature distribution, the vertex correction is generally non-zero and has to be implemented in the kinetic equation and self-energy according to the procedure described in Ref.~\cite{buchholdmethod}. As a result, the numerical computation becomes more involved but still feasible.
From the time evolved phonon occupation, we then determine the static structure factor $S^s_{\tau}(q)$ (Fig.~\ref{fig:SSFT}) and the scaling of the quasi-particle lifetimes $t_q$ (Fig.~\ref{fig:ScalingT}) as a function of time and momentum.

Our analysis of the phonon occupation dynamics for $T=0$ showed that dynamics is mainly governed by a fast energy redistribution for the high energy modes and a comparably slow evolution in the low frequency sector. This results from the structure of the particle number conserving vertex and the pinning of $n_{q=0}(\tau)=n_{q=0}(\tau=0)$ to its initial value. For a $T>0$ intial state, the situation does not change and the initial temperature has only little influence on the dynamics of $n_q$. The latter is dominated by the effect that drives the system away from equilibrium, i.e. the constant heating. We therefore decompose $n_q$ into an initial part and another one which builds up in time according to
\eq{EqFinT2}{
n_q(\tau)=n_q(\tau=0)+\delta n_q(\tau).}
 From our numerical simulations, we find that $\delta n_q(\tau)$ is almost independent of the initial temperature and for $T>0$ qualitatively equivalent to the $T=0$ scenario. While our results are obtained for the full simulation of the kinetic equation, we will use the above finding to explain the results.

To determine the static structure factor, we employ Eq.~\eqref{eq:staticstruct}, leading to
\begin{eqnarray}
S^s_{\tau}(q)&=&\tfrac{|q|K}{\pi}\left(2n_q(\tau=0)+2\delta n_q(\tau)+1\right)\nonumber\\
&\overset{\nu |q|\ll T}{=}&\tfrac{2TK}{\nu\pi}+\tfrac{|q|K}{\pi}\left(2\delta n_q(\tau)+1\right).\label{EqFinT3}
\end{eqnarray}
Compared to the $T=0$ case, this leads to a non-zero structure factor for $q=0$, ${S^s_{\tau}(q=0)=\frac{2TK}{\nu\pi}}$. For $q>0$, we find a $q^2$ scaling of the structure factor in the nonequilibrium regime as in the zero temperature case. For larger momenta $\nu |q|>T$, $n_q(\tau=0)$ vanishes and the finite and zero temperature structure factor are identical. In particular, the two different momentum regimes can still be characterized by the $T=0$ scaling properties of $S^s(q)$, see Fig.~\ref{fig:SSFT}.

In view of the scaling of the quasi-particle lifetimes $t_q$, and thereby the dynamic structure factor, our analysis fully confirms and corroborates the following expected scenario, cf. Fig.~\ref{fig:ScalingT}. Most importantly, an additional scaling regime for momenta much smaller than the temperature $\nu|q|\ll T$ occurs. In this regime, lifetimes scale with the thermal exponent $t_q\sim q^{-3/2}$. For momenta in the range $\tfrac{T}{\nu}<|q|<q_0$, a momentum region showing non-equilibrium scaling $t_q\sim q^{-5/3}$ is found, while for $|q|>q_0$, again thermal scaling behavior emerges. For $\tfrac{T}{\nu}\ll q_0$, the nonequilibrium scaling regime is sufficiently large to be resolved numerically and analytically. Such a situation is presented in Fig.~\ref{fig:ScalingT}. For this specific example $\tfrac{\nu q_0}{T}\approx 12$, while the nonequilibrium scaling regime has a momentum window ranging from $q\sub{min} \approx 0.35 q/q_0$ to $q\sub{max}\approx 2q/q_0$, i.e. $\tfrac{q\sub{max}}{q\sub{min}}\approx 5.7$. The initial temperatures used here are in reach for current experiments \cite{nagerl13, schmiedmayernjp13}, holding the promise to resolve the discussed non-equilibrium scaling regime in the lab. We finally note that while the different scaling regimes may not be very clearly visible in Fig.~\ref{fig:ScalingT}, using the data collapse procedure described in Fig.~\ref{fig:DSF} will allow to clearly discriminate the different scaling regimes.

\section{Conclusions}\label{sec:Conc}
 We have identified a new universal non-equilibrium scaling regime in the heating dynamics of interacting Luttinger liquids. Its existence is granted by the defining property of a Luttinger liquid, the presence of a sharp phonon mode, which prevails in the number conserving, permanently forward evolving open system. This triggers the hope that large classes of one-dimensional systems subject to non-equilibrium drive could exhibit similarly strong notions of universality as their equilibrium counterparts.  
Both extending the theoretical concepts familiar from closed systems, and exploring the status of universality in low dimensional driven open quantum systems, represent fascinating challenges for future research.

\begin{acknowledgements}
We thank L. Bonnes, M. Heyl, A. L\"auchli, J. Marino, H.-C. N\"agerl,  J. Schmiedmayer, L. Sieberer and P. Strack for useful discussions.  This research was supported
by the Austrian Science Fund (FWF) through the START grant Y 581-N16 and the SFB
FoQuS (FWF Project No. F4006-N16).
\end{acknowledgements}

\bibliography{bibliography}

\begin{thebibliography}{54}%
\makeatletter
\providecommand \@ifxundefined [1]{%
 \@ifx{#1\undefined}
}%
\providecommand \@ifnum [1]{%
 \ifnum #1\expandafter \@firstoftwo
 \else \expandafter \@secondoftwo
 \fi
}%
\providecommand \@ifx [1]{%
 \ifx #1\expandafter \@firstoftwo
 \else \expandafter \@secondoftwo
 \fi
}%
\providecommand \natexlab [1]{#1}%
\providecommand \enquote  [1]{``#1''}%
\providecommand \bibnamefont  [1]{#1}%
\providecommand \bibfnamefont [1]{#1}%
\providecommand \citenamefont [1]{#1}%
\providecommand \href@noop [0]{\@secondoftwo}%
\providecommand \href [0]{\begingroup \@sanitize@url \@href}%
\providecommand \@href[1]{\@@startlink{#1}\@@href}%
\providecommand \@@href[1]{\endgroup#1\@@endlink}%
\providecommand \@sanitize@url [0]{\catcode `\\12\catcode `\$12\catcode
  `\&12\catcode `\#12\catcode `\^12\catcode `\_12\catcode `\%12\relax}%
\providecommand \@@startlink[1]{}%
\providecommand \@@endlink[0]{}%
\providecommand \url  [0]{\begingroup\@sanitize@url \@url }%
\providecommand \@url [1]{\endgroup\@href {#1}{\urlprefix }}%
\providecommand \urlprefix  [0]{URL }%
\providecommand \Eprint [0]{\href }%
\providecommand \doibase [0]{http://dx.doi.org/}%
\providecommand \selectlanguage [0]{\@gobble}%
\providecommand \bibinfo  [0]{\@secondoftwo}%
\providecommand \bibfield  [0]{\@secondoftwo}%
\providecommand \translation [1]{[#1]}%
\providecommand \BibitemOpen [0]{}%
\providecommand \bibitemStop [0]{}%
\providecommand \bibitemNoStop [0]{.\EOS\space}%
\providecommand \EOS [0]{\spacefactor3000\relax}%
\providecommand \BibitemShut  [1]{\csname bibitem#1\endcsname}%
\let\auto@bib@innerbib\@empty
\bibitem [{\citenamefont {Haldane}(1981)}]{haldane81}%
  \BibitemOpen
  \bibfield  {author} {\bibinfo {author} {\bibfnamefont {F.~D.~M.}\
  \bibnamefont {Haldane}},\ }\href {\doibase 10.1103/PhysRevLett.47.1840}
  {\bibfield  {journal} {\bibinfo  {journal} {Phys. Rev. Lett.}\ }\textbf
  {\bibinfo {volume} {47}},\ \bibinfo {pages} {1840} (\bibinfo {year}
  {1981})}\BibitemShut {NoStop}%
\bibitem [{\citenamefont {Giamarchi}(2004)}]{giamarchibook}%
  \BibitemOpen
  \bibfield  {author} {\bibinfo {author} {\bibfnamefont {T.}~\bibnamefont
  {Giamarchi}},\ }\href@noop {} {\emph {\bibinfo {title} {Quantum Physics in
  One Dimension}}},\ International Series of Monographs on Physics\ (\bibinfo
  {publisher} {Oxford University Press},\ \bibinfo {address} {Oxford},\
  \bibinfo {year} {2004})\BibitemShut {NoStop}%
\bibitem [{\citenamefont {Andreev}(1980)}]{andreev80}%
  \BibitemOpen
  \bibfield  {author} {\bibinfo {author} {\bibfnamefont {A.~F.}\ \bibnamefont
  {Andreev}},\ }\href@noop {} {\bibfield  {journal} {\bibinfo  {journal} {Sov.
  Physics JETP}\ }\textbf {\bibinfo {volume} {51}},\ \bibinfo {pages} {1038}
  (\bibinfo {year} {1980})}\BibitemShut {NoStop}%
\bibitem [{\citenamefont {Samokhin}(1998)}]{samokhin98}%
  \BibitemOpen
  \bibfield  {author} {\bibinfo {author} {\bibfnamefont {K.~V.}\ \bibnamefont
  {Samokhin}},\ }\href@noop {} {\bibfield  {journal} {\bibinfo  {journal} {J.
  Phys. Condens. Matter}\ }\textbf {\bibinfo {volume} {10}},\ \bibinfo {pages}
  {L533} (\bibinfo {year} {1998})}\BibitemShut {NoStop}%
\bibitem [{\citenamefont {Narayan}\ and\ \citenamefont
  {Ramaswamy}(2002)}]{narayan02}%
  \BibitemOpen
  \bibfield  {author} {\bibinfo {author} {\bibfnamefont {O.}~\bibnamefont
  {Narayan}}\ and\ \bibinfo {author} {\bibfnamefont {S.}~\bibnamefont
  {Ramaswamy}},\ }\href {\doibase 10.1103/PhysRevLett.89.200601} {\bibfield
  {journal} {\bibinfo  {journal} {Phys. Rev. Lett.}\ }\textbf {\bibinfo
  {volume} {89}},\ \bibinfo {pages} {200601} (\bibinfo {year}
  {2002})}\BibitemShut {NoStop}%
\bibitem [{\citenamefont {Lepri}\ \emph {et~al.}(2003)\citenamefont {Lepri},
  \citenamefont {Livi},\ and\ \citenamefont {Politi}}]{lepri03}%
  \BibitemOpen
  \bibfield  {author} {\bibinfo {author} {\bibfnamefont {S.}~\bibnamefont
  {Lepri}}, \bibinfo {author} {\bibfnamefont {R.}~\bibnamefont {Livi}}, \ and\
  \bibinfo {author} {\bibfnamefont {A.}~\bibnamefont {Politi}},\ }\href
  {\doibase 10.1103/PhysRevE.68.067102} {\bibfield  {journal} {\bibinfo
  {journal} {Phys. Rev. E}\ }\textbf {\bibinfo {volume} {68}},\ \bibinfo
  {pages} {067102} (\bibinfo {year} {2003})}\BibitemShut {NoStop}%
\bibitem [{\citenamefont {Kardar}\ \emph {et~al.}(1986)\citenamefont {Kardar},
  \citenamefont {Parisi},\ and\ \citenamefont {Zhang}}]{KPZ}%
  \BibitemOpen
  \bibfield  {author} {\bibinfo {author} {\bibfnamefont {M.}~\bibnamefont
  {Kardar}}, \bibinfo {author} {\bibfnamefont {G.}~\bibnamefont {Parisi}}, \
  and\ \bibinfo {author} {\bibfnamefont {Y.-Z.}\ \bibnamefont {Zhang}},\
  }\href@noop {} {\bibfield  {journal} {\bibinfo  {journal} {Phys. Rev. Lett.}\
  }\textbf {\bibinfo {volume} {56}},\ \bibinfo {pages} {889} (\bibinfo {year}
  {1986})}\BibitemShut {NoStop}%
\bibitem [{\citenamefont {van Beijeren}(2012)}]{vanBeijeren}%
  \BibitemOpen
  \bibfield  {author} {\bibinfo {author} {\bibfnamefont {H.}~\bibnamefont {van
  Beijeren}},\ }\href@noop {} {\bibfield  {journal} {\bibinfo  {journal} {Phys.
  Rev. Lett.}\ }\textbf {\bibinfo {volume} {108}},\ \bibinfo {pages} {180601}
  (\bibinfo {year} {2012})}\BibitemShut {NoStop}%
\bibitem [{\citenamefont {Kulkarni}\ and\ \citenamefont
  {Lamacraft}(2013)}]{lamacraft13}%
  \BibitemOpen
  \bibfield  {author} {\bibinfo {author} {\bibfnamefont {M.}~\bibnamefont
  {Kulkarni}}\ and\ \bibinfo {author} {\bibfnamefont {A.}~\bibnamefont
  {Lamacraft}},\ }\href {\doibase 10.1103/PhysRevA.88.021603} {\bibfield
  {journal} {\bibinfo  {journal} {Phys. Rev. A}\ }\textbf {\bibinfo {volume}
  {88}},\ \bibinfo {pages} {021603} (\bibinfo {year} {2013})}\BibitemShut
  {NoStop}%
\bibitem [{\citenamefont {Arzamasovs}\ \emph {et~al.}(2013)\citenamefont
  {Arzamasovs}, \citenamefont {Bovo},\ and\ \citenamefont
  {Gangardt}}]{gangardt13}%
  \BibitemOpen
  \bibfield  {author} {\bibinfo {author} {\bibfnamefont {F.}~\bibnamefont
  {Arzamasovs}}, \bibinfo {author} {\bibfnamefont {F.}~\bibnamefont {Bovo}}, \
  and\ \bibinfo {author} {\bibfnamefont {D.}~\bibnamefont {Gangardt}},\
  }\href@noop {} {\bibfield  {journal} {\bibinfo  {journal} {arXiv: 1309.2647}\
  } (\bibinfo {year} {2013})}\BibitemShut {NoStop}%
\bibitem [{\citenamefont {Pereira}\ \emph {et~al.}(2006)\citenamefont
  {Pereira}, \citenamefont {Sirker}, \citenamefont {Caux}, \citenamefont
  {Hagemans}, \citenamefont {Maillet}, \citenamefont {White},\ and\
  \citenamefont {Affleck}}]{affleck06}%
  \BibitemOpen
  \bibfield  {author} {\bibinfo {author} {\bibfnamefont {R.~G.}\ \bibnamefont
  {Pereira}}, \bibinfo {author} {\bibfnamefont {J.}~\bibnamefont {Sirker}},
  \bibinfo {author} {\bibfnamefont {J.-S.}\ \bibnamefont {Caux}}, \bibinfo
  {author} {\bibfnamefont {R.}~\bibnamefont {Hagemans}}, \bibinfo {author}
  {\bibfnamefont {J.~M.}\ \bibnamefont {Maillet}}, \bibinfo {author}
  {\bibfnamefont {S.~R.}\ \bibnamefont {White}}, \ and\ \bibinfo {author}
  {\bibfnamefont {I.}~\bibnamefont {Affleck}},\ }\href {\doibase
  10.1103/PhysRevLett.96.257202} {\bibfield  {journal} {\bibinfo  {journal}
  {Phys. Rev. Lett.}\ }\textbf {\bibinfo {volume} {96}},\ \bibinfo {pages}
  {257202} (\bibinfo {year} {2006})}\BibitemShut {NoStop}%
\bibitem [{\citenamefont {Punk}\ and\ \citenamefont
  {Zwerger}(2006)}]{zwerger06}%
  \BibitemOpen
  \bibfield  {author} {\bibinfo {author} {\bibfnamefont {M.}~\bibnamefont
  {Punk}}\ and\ \bibinfo {author} {\bibfnamefont {W.}~\bibnamefont {Zwerger}},\
  }\href {\doibase 10.1088/1367-2630/8/8/168} {\bibfield  {journal} {\bibinfo
  {journal} {New J. Phys.}\ }\textbf {\bibinfo {volume} {8}},\ \bibinfo {pages}
  {168} (\bibinfo {year} {2006})}\BibitemShut {NoStop}%
\bibitem [{\citenamefont {Caux}\ and\ \citenamefont
  {Calabrese}(2006)}]{caux06}%
  \BibitemOpen
  \bibfield  {author} {\bibinfo {author} {\bibfnamefont {J.-S.}\ \bibnamefont
  {Caux}}\ and\ \bibinfo {author} {\bibfnamefont {P.}~\bibnamefont
  {Calabrese}},\ }\href@noop {} {\bibfield  {journal} {\bibinfo  {journal}
  {Phys. Rev. A}\ }\textbf {\bibinfo {volume} {74}},\ \bibinfo {pages} {031605}
  (\bibinfo {year} {2006})}\BibitemShut {NoStop}%
\bibitem [{\citenamefont {Gring}\ \emph {et~al.}(2012)\citenamefont {Gring},
  \citenamefont {Kuhnert}, \citenamefont {Langen}, \citenamefont {Kitagawa},
  \citenamefont {Rauer}, \citenamefont {Schreitl}, \citenamefont {Mazets},
  \citenamefont {Smith}, \citenamefont {Demler},\ and\ \citenamefont
  {Schmiedmayer}}]{schmiedmayer12}%
  \BibitemOpen
  \bibfield  {author} {\bibinfo {author} {\bibfnamefont {M.}~\bibnamefont
  {Gring}}, \bibinfo {author} {\bibfnamefont {M.}~\bibnamefont {Kuhnert}},
  \bibinfo {author} {\bibfnamefont {T.}~\bibnamefont {Langen}}, \bibinfo
  {author} {\bibfnamefont {T.}~\bibnamefont {Kitagawa}}, \bibinfo {author}
  {\bibfnamefont {B.}~\bibnamefont {Rauer}}, \bibinfo {author} {\bibfnamefont
  {M.}~\bibnamefont {Schreitl}}, \bibinfo {author} {\bibfnamefont
  {I.}~\bibnamefont {Mazets}}, \bibinfo {author} {\bibfnamefont {D.~A.}\
  \bibnamefont {Smith}}, \bibinfo {author} {\bibfnamefont {E.}~\bibnamefont
  {Demler}}, \ and\ \bibinfo {author} {\bibfnamefont {J.}~\bibnamefont
  {Schmiedmayer}},\ }\href {\doibase 10.1126/science.1224953} {\bibfield
  {journal} {\bibinfo  {journal} {Science}\ }\textbf {\bibinfo {volume}
  {337}},\ \bibinfo {pages} {1318} (\bibinfo {year} {2012})}\BibitemShut
  {NoStop}%
\bibitem [{\citenamefont {Langen}\ \emph {et~al.}(2013)\citenamefont {Langen},
  \citenamefont {Geiger}, \citenamefont {Kuhnert},\ and\ \citenamefont
  {Rauer}}]{schmiedmayernphys12}%
  \BibitemOpen
  \bibfield  {author} {\bibinfo {author} {\bibfnamefont {T.}~\bibnamefont
  {Langen}}, \bibinfo {author} {\bibfnamefont {R.}~\bibnamefont {Geiger}},
  \bibinfo {author} {\bibfnamefont {M.}~\bibnamefont {Kuhnert}}, \ and\
  \bibinfo {author} {\bibfnamefont {J.}~\bibnamefont {Rauer}, \bibfnamefont
  {B.~andSchmiedmayer}},\ }\href {\doibase 10.1038/nphys2739} {\bibfield
  {journal} {\bibinfo  {journal} {Nat. Phys.}\ }\textbf {\bibinfo {volume}
  {9}},\ \bibinfo {pages} {640} (\bibinfo {year} {2013})}\BibitemShut {NoStop}%
\bibitem [{\citenamefont {Trotzky}\ \emph {et~al.}(2013)\citenamefont
  {Trotzky}, \citenamefont {Chen}, \citenamefont {Flesch}, \citenamefont
  {McCulloch}, \citenamefont {Schollwock}, \citenamefont {Eisert},\ and\
  \citenamefont {Bloch}}]{bloch13}%
  \BibitemOpen
  \bibfield  {author} {\bibinfo {author} {\bibfnamefont {S.}~\bibnamefont
  {Trotzky}}, \bibinfo {author} {\bibfnamefont {Y.-A.}\ \bibnamefont {Chen}},
  \bibinfo {author} {\bibfnamefont {A.}~\bibnamefont {Flesch}}, \bibinfo
  {author} {\bibfnamefont {I.~P.}\ \bibnamefont {McCulloch}}, \bibinfo {author}
  {\bibfnamefont {U.}~\bibnamefont {Schollwock}}, \bibinfo {author}
  {\bibfnamefont {J.}~\bibnamefont {Eisert}}, \ and\ \bibinfo {author}
  {\bibfnamefont {I.}~\bibnamefont {Bloch}},\ }\href {\doibase
  10.1038/nphys2232} {\bibfield  {journal} {\bibinfo  {journal} {Nat. Phys.}\
  }\textbf {\bibinfo {volume} {8}},\ \bibinfo {pages} {325} (\bibinfo {year}
  {2013})}\BibitemShut {NoStop}%
\bibitem [{\citenamefont {Meinert}\ \emph {et~al.}(2013)\citenamefont
  {Meinert}, \citenamefont {Mark}, \citenamefont {Kirilov}, \citenamefont
  {Lauber}, \citenamefont {Weinmann}, \citenamefont {Daley},\ and\
  \citenamefont {N\"agerl}}]{nagerl13}%
  \BibitemOpen
  \bibfield  {author} {\bibinfo {author} {\bibfnamefont {F.}~\bibnamefont
  {Meinert}}, \bibinfo {author} {\bibfnamefont {M.~J.}\ \bibnamefont {Mark}},
  \bibinfo {author} {\bibfnamefont {E.}~\bibnamefont {Kirilov}}, \bibinfo
  {author} {\bibfnamefont {K.}~\bibnamefont {Lauber}}, \bibinfo {author}
  {\bibfnamefont {P.}~\bibnamefont {Weinmann}}, \bibinfo {author}
  {\bibfnamefont {A.~J.}\ \bibnamefont {Daley}}, \ and\ \bibinfo {author}
  {\bibfnamefont {H.-C.}\ \bibnamefont {N\"agerl}},\ }\href {\doibase
  10.1103/PhysRevLett.111.053003} {\bibfield  {journal} {\bibinfo  {journal}
  {Phys. Rev. Lett.}\ }\textbf {\bibinfo {volume} {111}},\ \bibinfo {pages}
  {053003} (\bibinfo {year} {2013})}\BibitemShut {NoStop}%
\bibitem [{\citenamefont {Berges}\ \emph {et~al.}(2004)\citenamefont {Berges},
  \citenamefont {Bors\'anyi},\ and\ \citenamefont
  {Wetterich}}]{berges_pretherm}%
  \BibitemOpen
  \bibfield  {author} {\bibinfo {author} {\bibfnamefont {J.}~\bibnamefont
  {Berges}}, \bibinfo {author} {\bibfnamefont {S.}~\bibnamefont {Bors\'anyi}},
  \ and\ \bibinfo {author} {\bibfnamefont {C.}~\bibnamefont {Wetterich}},\
  }\href {\doibase 10.1103/PhysRevLett.93.142002} {\bibfield  {journal}
  {\bibinfo  {journal} {Phys. Rev. Lett.}\ }\textbf {\bibinfo {volume} {93}},\
  \bibinfo {pages} {142002} (\bibinfo {year} {2004})}\BibitemShut {NoStop}%
\bibitem [{\citenamefont {Burkov}\ \emph {et~al.}(2007)\citenamefont {Burkov},
  \citenamefont {Lukin},\ and\ \citenamefont {Demler}}]{burkov07}%
  \BibitemOpen
  \bibfield  {author} {\bibinfo {author} {\bibfnamefont {A.~A.}\ \bibnamefont
  {Burkov}}, \bibinfo {author} {\bibfnamefont {M.~D.}\ \bibnamefont {Lukin}}, \
  and\ \bibinfo {author} {\bibfnamefont {E.}~\bibnamefont {Demler}},\ }\href
  {\doibase 10.1103/PhysRevLett.98.200404} {\bibfield  {journal} {\bibinfo
  {journal} {Phys. Rev. Lett.}\ }\textbf {\bibinfo {volume} {98}},\ \bibinfo
  {pages} {200404} (\bibinfo {year} {2007})}\BibitemShut {NoStop}%
\bibitem [{\citenamefont {Mitra}\ and\ \citenamefont
  {Giamarchi}(2011)}]{mitragiamarchi11}%
  \BibitemOpen
  \bibfield  {author} {\bibinfo {author} {\bibfnamefont {A.}~\bibnamefont
  {Mitra}}\ and\ \bibinfo {author} {\bibfnamefont {T.}~\bibnamefont
  {Giamarchi}},\ }\href {\doibase 10.1103/PhysRevLett.107.150602} {\bibfield
  {journal} {\bibinfo  {journal} {Phys. Rev. Lett.}\ }\textbf {\bibinfo
  {volume} {107}},\ \bibinfo {pages} {150602} (\bibinfo {year}
  {2011})}\BibitemShut {NoStop}%
\bibitem [{\citenamefont {Mitra}\ and\ \citenamefont
  {Giamarchi}(2012)}]{mitragiamarchi12}%
  \BibitemOpen
  \bibfield  {author} {\bibinfo {author} {\bibfnamefont {A.}~\bibnamefont
  {Mitra}}\ and\ \bibinfo {author} {\bibfnamefont {T.}~\bibnamefont
  {Giamarchi}},\ }\href {\doibase 10.1103/PhysRevB.85.075117} {\bibfield
  {journal} {\bibinfo  {journal} {Phys. Rev. B}\ }\textbf {\bibinfo {volume}
  {85}},\ \bibinfo {pages} {075117} (\bibinfo {year} {2012})}\BibitemShut
  {NoStop}%
\bibitem [{\citenamefont {Karrasch}\ \emph {et~al.}(2012)\citenamefont
  {Karrasch}, \citenamefont {Rentrop}, \citenamefont {Schuricht},\ and\
  \citenamefont {Meden}}]{Karrasch12}%
  \BibitemOpen
  \bibfield  {author} {\bibinfo {author} {\bibfnamefont {C.}~\bibnamefont
  {Karrasch}}, \bibinfo {author} {\bibfnamefont {J.}~\bibnamefont {Rentrop}},
  \bibinfo {author} {\bibfnamefont {D.}~\bibnamefont {Schuricht}}, \ and\
  \bibinfo {author} {\bibfnamefont {V.}~\bibnamefont {Meden}},\ }\href
  {\doibase 10.1103/PhysRevLett.109.126406} {\bibfield  {journal} {\bibinfo
  {journal} {Phys. Rev. Lett.}\ }\textbf {\bibinfo {volume} {109}},\ \bibinfo
  {pages} {126406} (\bibinfo {year} {2012})}\BibitemShut {NoStop}%
\bibitem [{\citenamefont {Tavora}\ and\ \citenamefont
  {Mitra}(2013)}]{Tavora13}%
  \BibitemOpen
  \bibfield  {author} {\bibinfo {author} {\bibfnamefont {M.}~\bibnamefont
  {Tavora}}\ and\ \bibinfo {author} {\bibfnamefont {A.}~\bibnamefont {Mitra}},\
  }\href@noop {} {\bibfield  {journal} {\bibinfo  {journal} {Phys. Rev. B}\
  }\textbf {\bibinfo {volume} {88}},\ \bibinfo {pages} {115144} (\bibinfo
  {year} {2013})}\BibitemShut {NoStop}%
\bibitem [{\citenamefont {Huber}\ and\ \citenamefont
  {Altman}(2009)}]{huberaltman11}%
  \BibitemOpen
  \bibfield  {author} {\bibinfo {author} {\bibfnamefont {S.~D.}\ \bibnamefont
  {Huber}}\ and\ \bibinfo {author} {\bibfnamefont {E.}~\bibnamefont {Altman}},\
  }\href {\doibase 10.1103/PhysRevLett.103.160402} {\bibfield  {journal}
  {\bibinfo  {journal} {Phys. Rev. Lett.}\ }\textbf {\bibinfo {volume} {103}},\
  \bibinfo {pages} {160402} (\bibinfo {year} {2009})}\BibitemShut {NoStop}%
\bibitem [{\citenamefont {Heyl}\ \emph {et~al.}(2013)\citenamefont {Heyl},
  \citenamefont {Polkovnikov},\ and\ \citenamefont {Kehrein}}]{heyl13}%
  \BibitemOpen
  \bibfield  {author} {\bibinfo {author} {\bibfnamefont {M.}~\bibnamefont
  {Heyl}}, \bibinfo {author} {\bibfnamefont {A.}~\bibnamefont {Polkovnikov}}, \
  and\ \bibinfo {author} {\bibfnamefont {S.}~\bibnamefont {Kehrein}},\ }\href
  {\doibase 10.1103/PhysRevLett.110.135704} {\bibfield  {journal} {\bibinfo
  {journal} {Phys. Rev. Lett.}\ }\textbf {\bibinfo {volume} {110}},\ \bibinfo
  {pages} {135704} (\bibinfo {year} {2013})}\BibitemShut {NoStop}%
\bibitem [{\citenamefont {Lux}\ \emph {et~al.}(2014)\citenamefont {Lux},
  \citenamefont {M{\"u}ller}, \citenamefont {Mitra},\ and\ \citenamefont
  {Rosch}}]{lux13}%
  \BibitemOpen
  \bibfield  {author} {\bibinfo {author} {\bibfnamefont {J.}~\bibnamefont
  {Lux}}, \bibinfo {author} {\bibfnamefont {J.}~\bibnamefont {M{\"u}ller}},
  \bibinfo {author} {\bibfnamefont {A.}~\bibnamefont {Mitra}}, \ and\ \bibinfo
  {author} {\bibfnamefont {A.}~\bibnamefont {Rosch}},\ }\href@noop {}
  {\bibfield  {journal} {\bibinfo  {journal} {Phys. Rev. A}\ }\textbf {\bibinfo
  {volume} {89}},\ \bibinfo {pages} {053608} (\bibinfo {year}
  {2014})}\BibitemShut {NoStop}%
\bibitem [{\citenamefont {Calabrese}\ and\ \citenamefont
  {Cardy}(2006)}]{cardycalabrese06}%
  \BibitemOpen
  \bibfield  {author} {\bibinfo {author} {\bibfnamefont {P.}~\bibnamefont
  {Calabrese}}\ and\ \bibinfo {author} {\bibfnamefont {J.}~\bibnamefont
  {Cardy}},\ }\href {\doibase 10.1103/PhysRevLett.96.136801} {\bibfield
  {journal} {\bibinfo  {journal} {Phys. Rev. Lett.}\ }\textbf {\bibinfo
  {volume} {96}},\ \bibinfo {pages} {136801} (\bibinfo {year}
  {2006})}\BibitemShut {NoStop}%
\bibitem [{\citenamefont {Marino}\ and\ \citenamefont
  {Silva}(2012)}]{Marino12}%
  \BibitemOpen
  \bibfield  {author} {\bibinfo {author} {\bibfnamefont {J.}~\bibnamefont
  {Marino}}\ and\ \bibinfo {author} {\bibfnamefont {A.}~\bibnamefont {Silva}},\
  }\href@noop {} {\bibfield  {journal} {\bibinfo  {journal} {Phys. Rev. B}\
  }\textbf {\bibinfo {volume} {86}},\ \bibinfo {pages} {060408} (\bibinfo
  {year} {2012})}\BibitemShut {NoStop}%
\bibitem [{\citenamefont {Schachenmayer}\ \emph {et~al.}(2014)\citenamefont
  {Schachenmayer}, \citenamefont {Pollet}, \citenamefont {Troyer},\ and\
  \citenamefont {Daley}}]{schachen14}%
  \BibitemOpen
  \bibfield  {author} {\bibinfo {author} {\bibfnamefont {J.}~\bibnamefont
  {Schachenmayer}}, \bibinfo {author} {\bibfnamefont {L.}~\bibnamefont
  {Pollet}}, \bibinfo {author} {\bibfnamefont {M.}~\bibnamefont {Troyer}}, \
  and\ \bibinfo {author} {\bibfnamefont {A.~J.}\ \bibnamefont {Daley}},\ }\href
  {http://link.aps.org/doi/10.1103/PhysRevA.89.011601} {\bibfield  {journal}
  {\bibinfo  {journal} {Phys. Rev. A}\ }\textbf {\bibinfo {volume} {89}},\
  \bibinfo {pages} {011601} (\bibinfo {year} {2014})}\BibitemShut {NoStop}%
\bibitem [{\citenamefont {Pichler}\ \emph {et~al.}(2010)\citenamefont
  {Pichler}, \citenamefont {Daley},\ and\ \citenamefont {Zoller}}]{pichler10}%
  \BibitemOpen
  \bibfield  {author} {\bibinfo {author} {\bibfnamefont {H.}~\bibnamefont
  {Pichler}}, \bibinfo {author} {\bibfnamefont {A.~J.}\ \bibnamefont {Daley}},
  \ and\ \bibinfo {author} {\bibfnamefont {P.}~\bibnamefont {Zoller}},\ }\href
  {\doibase 10.1103/PhysRevA.82.063605} {\bibfield  {journal} {\bibinfo
  {journal} {Phys. Rev. A}\ }\textbf {\bibinfo {volume} {82}},\ \bibinfo
  {pages} {063605} (\bibinfo {year} {2010})}\BibitemShut {NoStop}%
\bibitem [{Note1()}]{Note1}%
  \BibitemOpen
  \bibinfo {note} {The effective mass $m^{-1}=\partial _q^2\epsilon _q$ in the
  lattice evaluates to $m=(4Ja^2)^{-1}$, $a$ being the lattice
  constant.}\BibitemShut {Stop}%
\bibitem [{Note2()}]{Note2}%
  \BibitemOpen
  \bibinfo {note} {The effective heating rate is defined as ${\gamma _{\unhbox
  \voidb@x \hbox {\relax \protect \fontsize {5}{6}\protect \selectfont
  H}}=\gamma _{\unhbox \voidb@x \hbox {\relax \protect \fontsize {5}{6}\protect
  \selectfont E}}\left (\DOTSB \sum@ \slimits@ _{q<\Lambda }2\nu q^2\right
  )^{-1}}$.}\BibitemShut {Stop}%
\bibitem [{Note3()}]{Note3}%
  \BibitemOpen
  \bibinfo {note} {A term $\sim (\partial _x \phi )^3 $, not present in the
  microscopic theory, only slightly modifies prefactors, but not the scaling
  laws, while a contribution $\sim (\partial _x \theta )^3$ is ruled out by the
  $\theta \to -\theta $ symmetry of the Hamiltonian.}\BibitemShut {Stop}%
\bibitem [{\citenamefont {Rohringer}\ \emph {et~al.}(2013)\citenamefont
  {Rohringer}, \citenamefont {Fischer}, \citenamefont {Steiner}, \citenamefont
  {Mazets}, \citenamefont {Schmiedmayer},\ and\ \citenamefont
  {Trupke}}]{trupke13}%
  \BibitemOpen
  \bibfield  {author} {\bibinfo {author} {\bibfnamefont {W.}~\bibnamefont
  {Rohringer}}, \bibinfo {author} {\bibfnamefont {D.}~\bibnamefont {Fischer}},
  \bibinfo {author} {\bibfnamefont {F.}~\bibnamefont {Steiner}}, \bibinfo
  {author} {\bibfnamefont {I.~E.}\ \bibnamefont {Mazets}}, \bibinfo {author}
  {\bibfnamefont {J.}~\bibnamefont {Schmiedmayer}}, \ and\ \bibinfo {author}
  {\bibfnamefont {M.}~\bibnamefont {Trupke}},\ }\href@noop {} {\bibfield
  {journal} {\bibinfo  {journal} {arXiv: 1312.5948}\ } (\bibinfo {year}
  {2013})}\BibitemShut {NoStop}%
\bibitem [{Note4()}]{Note4}%
  \BibitemOpen
  \bibinfo {note} {{\protect \color {black}While the present heating mechanism
  corresponds to a fluctuating chemical potential, i.e. to the class of
  Langevin equations with additive noise, it is also possible to imagine
  mechanisms with multiplicative noise, such as fluctuating interaction
  parameters. However, while these would correspond to a completely different
  universality class, typically featuring an exponential energy increase, which
  is not relevant for optical lattice experiments, where the energy increase is
  linear in time.}}\BibitemShut {Stop}%
\bibitem [{\citenamefont {Poletti}\ \emph {et~al.}(2012)\citenamefont
  {Poletti}, \citenamefont {Bernier}, \citenamefont {Georges},\ and\
  \citenamefont {Kollath}}]{poletti12}%
  \BibitemOpen
  \bibfield  {author} {\bibinfo {author} {\bibfnamefont {D.}~\bibnamefont
  {Poletti}}, \bibinfo {author} {\bibfnamefont {J.-S.}\ \bibnamefont
  {Bernier}}, \bibinfo {author} {\bibfnamefont {A.}~\bibnamefont {Georges}}, \
  and\ \bibinfo {author} {\bibfnamefont {C.}~\bibnamefont {Kollath}},\ }\href
  {\doibase 10.1103/PhysRevLett.109.045302} {\bibfield  {journal} {\bibinfo
  {journal} {Phys. Rev. Lett.}\ }\textbf {\bibinfo {volume} {109}},\ \bibinfo
  {pages} {045302} (\bibinfo {year} {2012})}\BibitemShut {NoStop}%
\bibitem [{\citenamefont {Poletti}\ \emph {et~al.}(2013)\citenamefont
  {Poletti}, \citenamefont {Barmettler}, \citenamefont {Georges},\ and\
  \citenamefont {Kollath}}]{poletti13}%
  \BibitemOpen
  \bibfield  {author} {\bibinfo {author} {\bibfnamefont {D.}~\bibnamefont
  {Poletti}}, \bibinfo {author} {\bibfnamefont {P.}~\bibnamefont {Barmettler}},
  \bibinfo {author} {\bibfnamefont {A.}~\bibnamefont {Georges}}, \ and\
  \bibinfo {author} {\bibfnamefont {C.}~\bibnamefont {Kollath}},\ }\href
  {\doibase 10.1103/PhysRevLett.111.195301} {\bibfield  {journal} {\bibinfo
  {journal} {Phys. Rev. Lett.}\ }\textbf {\bibinfo {volume} {111}},\ \bibinfo
  {pages} {195301} (\bibinfo {year} {2013})}\BibitemShut {NoStop}%
\bibitem [{\citenamefont {Cai}\ and\ \citenamefont
  {Barthel}(2013)}]{caibarthel13}%
  \BibitemOpen
  \bibfield  {author} {\bibinfo {author} {\bibfnamefont {Z.}~\bibnamefont
  {Cai}}\ and\ \bibinfo {author} {\bibfnamefont {T.}~\bibnamefont {Barthel}},\
  }\href {\doibase 10.1103/PhysRevLett.111.150403} {\bibfield  {journal}
  {\bibinfo  {journal} {Phys. Rev. Lett.}\ }\textbf {\bibinfo {volume} {111}},\
  \bibinfo {pages} {150403} (\bibinfo {year} {2013})}\BibitemShut {NoStop}%
\bibitem [{\citenamefont {Altman}\ \emph {et~al.}(2015)\citenamefont {Altman},
  \citenamefont {Sieberer}, \citenamefont {Chen}, \citenamefont {Diehl},\ and\
  \citenamefont {Toner}}]{altman13}%
  \BibitemOpen
  \bibfield  {author} {\bibinfo {author} {\bibfnamefont {E.}~\bibnamefont
  {Altman}}, \bibinfo {author} {\bibfnamefont {L.~M.}\ \bibnamefont
  {Sieberer}}, \bibinfo {author} {\bibfnamefont {L.}~\bibnamefont {Chen}},
  \bibinfo {author} {\bibfnamefont {S.}~\bibnamefont {Diehl}}, \ and\ \bibinfo
  {author} {\bibfnamefont {J.}~\bibnamefont {Toner}},\ }\href@noop {}
  {\bibfield  {journal} {\bibinfo  {journal} {Phys. Rev. X}\ }\textbf {\bibinfo
  {volume} {5}},\ \bibinfo {pages} {011017} (\bibinfo {year}
  {2015})}\BibitemShut {NoStop}%
\bibitem [{\citenamefont {Kessler}(2012)}]{kessler12}%
  \BibitemOpen
  \bibfield  {author} {\bibinfo {author} {\bibfnamefont {E.~M.}\ \bibnamefont
  {Kessler}},\ }\href {\doibase 10.1103/PhysRevA.86.012126} {\bibfield
  {journal} {\bibinfo  {journal} {Phys. Rev. A}\ }\textbf {\bibinfo {volume}
  {86}},\ \bibinfo {pages} {012126} (\bibinfo {year} {2012})}\BibitemShut
  {NoStop}%
\bibitem [{\citenamefont {Degenfeld-Schonburg}\ and\ \citenamefont
  {Hartmann}(2014)}]{hartmann13}%
  \BibitemOpen
  \bibfield  {author} {\bibinfo {author} {\bibfnamefont {P.}~\bibnamefont
  {Degenfeld-Schonburg}}\ and\ \bibinfo {author} {\bibfnamefont {M.~J.}\
  \bibnamefont {Hartmann}},\ }\href@noop {} {\bibfield  {journal} {\bibinfo
  {journal} {Phys. Rev. B}\ }\textbf {\bibinfo {volume} {89}},\ \bibinfo
  {pages} {245108} (\bibinfo {year} {2014})}\BibitemShut {NoStop}%
\bibitem [{\citenamefont {Li}\ \emph {et~al.}(2013)\citenamefont {Li},
  \citenamefont {Petruccione},\ and\ \citenamefont {Koch}}]{koch13}%
  \BibitemOpen
  \bibfield  {author} {\bibinfo {author} {\bibfnamefont {A.}~\bibnamefont
  {Li}}, \bibinfo {author} {\bibfnamefont {F.}~\bibnamefont {Petruccione}}, \
  and\ \bibinfo {author} {\bibfnamefont {J.}~\bibnamefont {Koch}},\ }\href@noop
  {} {\bibfield  {journal} {\bibinfo  {journal} {arXiv: 1311.3227}\ } (\bibinfo
  {year} {2013})}\BibitemShut {NoStop}%
\bibitem [{\citenamefont {Torre}\ \emph {et~al.}(2013)\citenamefont {Torre},
  \citenamefont {Diehl}, \citenamefont {Lukin}, \citenamefont {Sachdev},\ and\
  \citenamefont {Strack}}]{dalla13}%
  \BibitemOpen
  \bibfield  {author} {\bibinfo {author} {\bibfnamefont {E.~G.~D.}\
  \bibnamefont {Torre}}, \bibinfo {author} {\bibfnamefont {S.}~\bibnamefont
  {Diehl}}, \bibinfo {author} {\bibfnamefont {M.~D.}\ \bibnamefont {Lukin}},
  \bibinfo {author} {\bibfnamefont {S.}~\bibnamefont {Sachdev}}, \ and\
  \bibinfo {author} {\bibfnamefont {P.}~\bibnamefont {Strack}},\ }\href
  {\doibase 10.1103/PhysRevA.87.023831} {\bibfield  {journal} {\bibinfo
  {journal} {Phys. Rev. A}\ }\textbf {\bibinfo {volume} {87}},\ \bibinfo
  {pages} {023831} (\bibinfo {year} {2013})}\BibitemShut {NoStop}%
\bibitem [{\citenamefont {Sieberer}\ \emph {et~al.}(2013)\citenamefont
  {Sieberer}, \citenamefont {Huber}, \citenamefont {Altman},\ and\
  \citenamefont
  {Diehl}}]{sieberer13:_dynam_critic_phenom_driven_dissip_system}%
  \BibitemOpen
  \bibfield  {author} {\bibinfo {author} {\bibfnamefont {L.~M.}\ \bibnamefont
  {Sieberer}}, \bibinfo {author} {\bibfnamefont {S.~D.}\ \bibnamefont {Huber}},
  \bibinfo {author} {\bibfnamefont {E.}~\bibnamefont {Altman}}, \ and\ \bibinfo
  {author} {\bibfnamefont {S.}~\bibnamefont {Diehl}},\ }\href {\doibase
  10.1103/PhysRevLett.110.195301} {\bibfield  {journal} {\bibinfo  {journal}
  {Phys. Rev. Lett.}\ }\textbf {\bibinfo {volume} {110}},\ \bibinfo {pages}
  {195301} (\bibinfo {year} {2013})}\BibitemShut {NoStop}%
\bibitem [{\citenamefont {Altland}\ and\ \citenamefont
  {Simons}(2010)}]{Altland/Simons}%
  \BibitemOpen
  \bibfield  {author} {\bibinfo {author} {\bibfnamefont {A.}~\bibnamefont
  {Altland}}\ and\ \bibinfo {author} {\bibfnamefont {B.}~\bibnamefont
  {Simons}},\ }\href@noop {} {\emph {\bibinfo {title} {Condensed Matter Field
  Theory}}},\ \bibinfo {edition} {2nd}\ ed.\ (\bibinfo  {publisher} {Cambridge
  University Press},\ \bibinfo {address} {Cambridge},\ \bibinfo {year}
  {2010})\BibitemShut {NoStop}%
\bibitem [{\citenamefont {Kamenev}(2011)}]{kamenevbook}%
  \BibitemOpen
  \bibfield  {author} {\bibinfo {author} {\bibfnamefont {A.}~\bibnamefont
  {Kamenev}},\ }\href@noop {} {\emph {\bibinfo {title} {Field Theory of
  Non-Equilibrium Systems}}}\ (\bibinfo  {publisher} {Cambridge University
  Press},\ \bibinfo {year} {2011})\BibitemShut {NoStop}%
\bibitem [{\citenamefont {Sieberer}\ \emph {et~al.}(2014)\citenamefont
  {Sieberer}, \citenamefont {Huber}, \citenamefont {Altman},\ and\
  \citenamefont {Diehl}}]{siebererlong13}%
  \BibitemOpen
  \bibfield  {author} {\bibinfo {author} {\bibfnamefont {L.~M.}\ \bibnamefont
  {Sieberer}}, \bibinfo {author} {\bibfnamefont {S.~D.}\ \bibnamefont {Huber}},
  \bibinfo {author} {\bibfnamefont {E.}~\bibnamefont {Altman}}, \ and\ \bibinfo
  {author} {\bibfnamefont {S.}~\bibnamefont {Diehl}},\ }\href@noop {}
  {\bibfield  {journal} {\bibinfo  {journal} {Phys. Rev. B}\ }\textbf {\bibinfo
  {volume} {89}},\ \bibinfo {pages} {134310} (\bibinfo {year}
  {2014})}\BibitemShut {NoStop}%
\bibitem [{\citenamefont {Buchhold}\ and\ \citenamefont
  {Diehl}(2015)}]{buchholdmethod}%
  \BibitemOpen
  \bibfield  {author} {\bibinfo {author} {\bibfnamefont {M.}~\bibnamefont
  {Buchhold}}\ and\ \bibinfo {author} {\bibfnamefont {S.}~\bibnamefont
  {Diehl}},\ }\href@noop {} {\bibfield  {journal} {\bibinfo  {journal} {arXiv:
  1501.01027}\ } (\bibinfo {year} {2015})}\BibitemShut {NoStop}%
\bibitem [{\citenamefont {Mahan}(1990)}]{mahan}%
  \BibitemOpen
  \bibfield  {author} {\bibinfo {author} {\bibfnamefont {G.~D.}\ \bibnamefont
  {Mahan}},\ }\href@noop {} {\emph {\bibinfo {title} {Many-Particle
  Physics}}},\ \bibinfo {edition} {2nd}\ ed.\ (\bibinfo  {publisher} {Plenum
  Press, New York},\ \bibinfo {year} {1990})\BibitemShut {NoStop}%
\bibitem [{\citenamefont {Stamper-Kurn}\ \emph {et~al.}(1999)\citenamefont
  {Stamper-Kurn}, \citenamefont {Chikkatur}, \citenamefont {G\"orlitz},
  \citenamefont {Inouye}, \citenamefont {Gupta}, \citenamefont {Pritchard},\
  and\ \citenamefont {Ketterle}}]{ketterle99}%
  \BibitemOpen
  \bibfield  {author} {\bibinfo {author} {\bibfnamefont {D.~M.}\ \bibnamefont
  {Stamper-Kurn}}, \bibinfo {author} {\bibfnamefont {A.~P.}\ \bibnamefont
  {Chikkatur}}, \bibinfo {author} {\bibfnamefont {A.}~\bibnamefont
  {G\"orlitz}}, \bibinfo {author} {\bibfnamefont {S.}~\bibnamefont {Inouye}},
  \bibinfo {author} {\bibfnamefont {S.}~\bibnamefont {Gupta}}, \bibinfo
  {author} {\bibfnamefont {D.~E.}\ \bibnamefont {Pritchard}}, \ and\ \bibinfo
  {author} {\bibfnamefont {W.}~\bibnamefont {Ketterle}},\ }\href {\doibase
  10.1103/PhysRevLett.83.2876} {\bibfield  {journal} {\bibinfo  {journal}
  {Phys. Rev. Lett.}\ }\textbf {\bibinfo {volume} {83}},\ \bibinfo {pages}
  {2876} (\bibinfo {year} {1999})}\BibitemShut {NoStop}%
\bibitem [{\citenamefont {Stenger}\ \emph {et~al.}(1999)\citenamefont
  {Stenger}, \citenamefont {Inouye}, \citenamefont {Chikkatur}, \citenamefont
  {Stamper-Kurn}, \citenamefont {Pritchard},\ and\ \citenamefont
  {Ketterle}}]{ketterleI99}%
  \BibitemOpen
  \bibfield  {author} {\bibinfo {author} {\bibfnamefont {J.}~\bibnamefont
  {Stenger}}, \bibinfo {author} {\bibfnamefont {S.}~\bibnamefont {Inouye}},
  \bibinfo {author} {\bibfnamefont {A.~P.}\ \bibnamefont {Chikkatur}}, \bibinfo
  {author} {\bibfnamefont {D.~M.}\ \bibnamefont {Stamper-Kurn}}, \bibinfo
  {author} {\bibfnamefont {D.~E.}\ \bibnamefont {Pritchard}}, \ and\ \bibinfo
  {author} {\bibfnamefont {W.}~\bibnamefont {Ketterle}},\ }\href {\doibase
  10.1103/PhysRevLett.82.4569} {\bibfield  {journal} {\bibinfo  {journal}
  {Phys. Rev. Lett.}\ }\textbf {\bibinfo {volume} {82}},\ \bibinfo {pages}
  {4569} (\bibinfo {year} {1999})}\BibitemShut {NoStop}%
\bibitem [{\citenamefont {St\"oferle}\ \emph {et~al.}(2004)\citenamefont
  {St\"oferle}, \citenamefont {Moritz}, \citenamefont {Schori}, \citenamefont
  {K\"ohl},\ and\ \citenamefont {Esslinger}}]{esslinger04}%
  \BibitemOpen
  \bibfield  {author} {\bibinfo {author} {\bibfnamefont {T.}~\bibnamefont
  {St\"oferle}}, \bibinfo {author} {\bibfnamefont {H.}~\bibnamefont {Moritz}},
  \bibinfo {author} {\bibfnamefont {C.}~\bibnamefont {Schori}}, \bibinfo
  {author} {\bibfnamefont {M.}~\bibnamefont {K\"ohl}}, \ and\ \bibinfo {author}
  {\bibfnamefont {T.}~\bibnamefont {Esslinger}},\ }\href {\doibase
  10.1103/PhysRevLett.92.130403} {\bibfield  {journal} {\bibinfo  {journal}
  {Phys. Rev. Lett.}\ }\textbf {\bibinfo {volume} {92}},\ \bibinfo {pages}
  {130403} (\bibinfo {year} {2004})}\BibitemShut {NoStop}%
\bibitem [{\citenamefont {Bissbort}\ \emph {et~al.}(2011)\citenamefont
  {Bissbort}, \citenamefont {G\"otze}, \citenamefont {Li}, \citenamefont
  {Heinze}, \citenamefont {Krauser}, \citenamefont {Weinberg}, \citenamefont
  {Becker}, \citenamefont {Sengstock},\ and\ \citenamefont
  {Hofstetter}}]{sengstock11}%
  \BibitemOpen
  \bibfield  {author} {\bibinfo {author} {\bibfnamefont {U.}~\bibnamefont
  {Bissbort}}, \bibinfo {author} {\bibfnamefont {S.}~\bibnamefont {G\"otze}},
  \bibinfo {author} {\bibfnamefont {Y.}~\bibnamefont {Li}}, \bibinfo {author}
  {\bibfnamefont {J.}~\bibnamefont {Heinze}}, \bibinfo {author} {\bibfnamefont
  {J.~S.}\ \bibnamefont {Krauser}}, \bibinfo {author} {\bibfnamefont
  {M.}~\bibnamefont {Weinberg}}, \bibinfo {author} {\bibfnamefont
  {C.}~\bibnamefont {Becker}}, \bibinfo {author} {\bibfnamefont
  {K.}~\bibnamefont {Sengstock}}, \ and\ \bibinfo {author} {\bibfnamefont
  {W.}~\bibnamefont {Hofstetter}},\ }\href
  {http://link.aps.org/doi/10.1103/PhysRevLett.106.205303} {\bibfield
  {journal} {\bibinfo  {journal} {Phys. Rev. Lett.}\ }\textbf {\bibinfo
  {volume} {106}},\ \bibinfo {pages} {205303} (\bibinfo {year}
  {2011})}\BibitemShut {NoStop}%
\bibitem [{\citenamefont {Smith}\ \emph {et~al.}(2013)\citenamefont {Smith},
  \citenamefont {Gring}, \citenamefont {Langen}, \citenamefont {Kuhnert},
  \citenamefont {Rauer}, \citenamefont {Geiger}, \citenamefont {Kitagawa},
  \citenamefont {Mazets}, \citenamefont {E.},\ and\ \citenamefont
  {Schmiedmayer}}]{schmiedmayernjp13}%
  \BibitemOpen
  \bibfield  {author} {\bibinfo {author} {\bibfnamefont {A.~D.}\ \bibnamefont
  {Smith}}, \bibinfo {author} {\bibfnamefont {M.}~\bibnamefont {Gring}},
  \bibinfo {author} {\bibfnamefont {T.}~\bibnamefont {Langen}}, \bibinfo
  {author} {\bibfnamefont {M.}~\bibnamefont {Kuhnert}}, \bibinfo {author}
  {\bibfnamefont {B.}~\bibnamefont {Rauer}}, \bibinfo {author} {\bibfnamefont
  {R.}~\bibnamefont {Geiger}}, \bibinfo {author} {\bibfnamefont
  {T.}~\bibnamefont {Kitagawa}}, \bibinfo {author} {\bibfnamefont
  {I.}~\bibnamefont {Mazets}}, \bibinfo {author} {\bibfnamefont
  {D.}~\bibnamefont {E.}}, \ and\ \bibinfo {author} {\bibfnamefont
  {J.}~\bibnamefont {Schmiedmayer}},\ }\href {\doibase
  10.1088/1367-2630/15/7/075011} {\bibfield  {journal} {\bibinfo  {journal}
  {New J. Phys.}\ }\textbf {\bibinfo {volume} {15}},\ \bibinfo {pages} {075011}
  (\bibinfo {year} {2013})}\BibitemShut {NoStop}%
\end{thebibliography}%

\end{document}